\documentclass[a4paper,11pt]{article}
\usepackage{jheppub} 
\usepackage{xcolor}
\usepackage{physics}
\usepackage{xspace}
\usepackage{mathrsfs}
\usepackage{soul}
\usepackage{mathtools}
\usepackage{tensor}
\usepackage{tabularx}
\usepackage{cancel}
\usepackage{booktabs}

\makeatletter
\g@addto@macro\bfseries{\boldmath}
\makeatother

\newcommand{\A}{\mathcal{A}}
\newcommand{\B}{\mathcal{B}}
\newcommand{\C}{\mathcal{C}}
\newcommand{\D}{\mathcal{D}}
\newcommand{\M}{\mathcal{M}}

\preprint{Imperial-TP-2024-CH-6\\ \rightline{UUITP-27/24}}
\title{Gauging generalised symmetries in linear gravity}

\author[a]{Chris Hull,}
\author[a]{Maxwell L. Hutt,}
\author[a,b]{Ulf Lindstr\"{o}m}

\affiliation[a]{The Blackett Laboratory, Imperial College London, Prince Consort Road, London, SW7 2AZ, UK}
\affiliation[b]{Department of Physics and Astronomy, Uppsala University,
Box 516, SE-75120 Uppsala, Sweden
and Centre for Geometry and Physics, Uppsala University,
Box 480, SE-75106 Uppsala, Sweden
}

\emailAdd{c.hull@imperial.ac.uk, m.hutt22@imperial.ac.uk, ulf.lindstrom@physics.uu.se}

\abstract{
The theory of a free spin-2 field on Minkowski spacetime  has  1-form and $(d-3)$-form symmetries associated with conserved currents formed by  contractions of the linearised Riemann tensor with conformal Killing-Yano 2-forms. We show that a subset of these can be interpreted as Noether currents for specific shift symmetries of the graviton that involve a Killing vector and a closed 1-form parameter. We give a systematic method to gauge these 1-form  symmetries by coupling the currents to background gauge fields and introducing a particular set of counter-terms involving the background fields. The simultaneous gauging of certain pairs of 1-form and $(d-3)$-form symmetries is obstructed by the presence of mixed 't Hooft anomalies. The anomalous pairs of symmetries are those which are related by gravitational duality. The implications of these anomalies are discussed.
}

\begin{document}
\maketitle
\flushbottom

\section{Introduction}
\label{sec:intro}

Gauge theories can often be formulated in several different ways, related by dualities. 
While different formulations of a theory may be written in terms of different fundamental fields with different gauge symmetries, the (generalised) \emph{global} symmetries must match in all formulations. These global symmetries can have 't Hooft anomalies (which prevent their gauging), and these must also then match between different formulations.
Because of their invariance under renormalisation group (RG) flow \cite{tHooft:1979rat}, these 't Hooft anomalies must be present at all energy scales and so carry important physical information about the theory.

A well-known example is that of $d$-dimensional pure Maxwell theory.\footnote{See, e.g. \cite{Gaiotto2015GeneralizedSymmetries, Hofman:2017vwr} and references therein for review of the higher-form symmetries of Maxwell theory.} This has both a 1-form and $(d-3)$-form symmetry. The 1-form symmetry corresponds to a 2-form current $J^{(1)} = F$, while the $(d-3)$-form symmetry corresponds to a $(d-2)$-form current $J^{(d-3)} = \star F$. These can be integrated to give charges
\begin{equation}
    Q^{(1)} = \int_{\Sigma_{d-2}} \star J^{(1)} \qc Q^{(d-3)} = \int_{\Sigma_2} \star J^{(d-3)}
\end{equation}
where $\Sigma_{d-2}$ and $\Sigma_2$ are a $(d-2)$-cycle and 2-cycle respectively.
In the standard formulation of the theory   in terms of a 1-form $U(1)$ connection $A$  the electric charge $Q^{(1)}$ is a Noether charge associated with a shift symmetry of $A$, whereas the magnetic charge $Q^{(d-3)}$ is a topological charge that is not associated with any field transformation. That is, the $(d-3)$-form symmetry is not a symmetry in the `traditional' sense but is rather a \emph{generalised symmetry} \cite{Gaiotto2015GeneralizedSymmetries}. As a consequence, $J^{(1)}$ is conserved only on-shell, whereas $J^{(d-3)}$ is conserved identically. For general gauge theories, we shall refer to a symmetry related to a field transformation, whose current is only conserved on-shell, as \emph{electric}, and a symmetry whose current is identically conserved as \emph{magnetic}.

There is a dual formulation written in terms of a $(d-3)$-form gauge field $\tilde{A}$. In this presentation, $Q^{(d-3)}$ is a Noether charge and is associated with a shift of $\tilde{A}$, while $Q^{(1)}$ is not related to any field transformation. When written in terms of $\tilde{A}$, it is $J^{(d-3)}$ which is conserved on-shell and $J^{(1)}$ which is identically conserved.
Therefore, in this presentation, $Q^{(d-3)}$ is an electric charge for $\tilde{A}$ and $Q^{(1)}$ is a magnetic charge.
The 1-form symmetry can be gauged by coupling $J^{(1)}$ to a 2-form background gauge field, and the $(d-3)$-form symmetry can be gauged by coupling $J^{(d-3)}$ to a $(d-2)$-form background gauge field. 
However, it is not possible to simultaneously couple these background fields such that the resulting action is gauge invariant under all background gauge transformations.
This is the statement that there is a mixed 't Hooft anomaly between the global 1-form and $(d-3)$-form symmetries. This structure can be generalised to $p$-form gauge theories, which play a central role in string theory and supergravity.

The theory of Einstein gravity linearised around Minkowski space has a similar symmetry structure that has received considerable attention recently  \cite{Hinterbichler2023GravitySymmetries, Benedetti2022GeneralizedGraviton, Benedetti2023GeneralizedGravitons, BenedettiNoether, Gomez-Fayren2023CovariantRelativity, Hull:2024mfb, Hull:2024xgo, Hull:2024bcl}. This symmetry structure generalises that of $p$-form gauge fields in several interesting ways. The standard formulation of the theory is written in terms of a spin-2 gauge field $h_{\mu\nu}$. In $d$ dimensions, this theory has a set of 1-form and $(d-3)$-form symmetries which can be interpreted as `electric' and `magnetic' respectively. In this work, we study the gauging of these symmetries and find the full set of mixed 't Hooft anomalies between them. 

The 1-form symmetries of the graviton are related to a set of conserved 2-forms
\begin{equation}
    Y_+(K)_{\mu\nu} = R_{\mu\nu\rho\sigma}K^{\rho\sigma} + 4R\indices{^\rho_{[\mu}} K_{\nu]\rho} + RK_{\mu\nu}
\end{equation}
where $R$ is the linearised Riemann tensor and $K$ is a conformal Killing-Yano (CKY) 2-form, which will be discussed in detail in section \ref{sec:Penrose_review}.
These 2-forms are close analogues to a set of conserved 2-forms studied by Penrose \cite{Penrose1982Quasi-localRelativity} and have been referred to as the \emph{improved Penrose 2-forms} in \cite{Hull:2024xgo}. They can be integrated to give the \emph{Penrose charges}
\begin{equation}
    Q(K) = \int_{\Sigma_{d-2}} \star Y_+(K)
\end{equation}
where $\Sigma_{d-2}$ is a codimension-2 cycle.
These charges generate a 1-form symmetry for each independent CKY 2-form $K$.

The graviton higher-form symmetries have  properties in $d=4$ dimensions that are different from those  in $d>4$.
In $d=4$ dimensions, the charges $Q(K)$ give rise to 1-form symmetries for all CKY 2-forms $K$. A subset of them are identically conserved and cannot be related to field transformations of the graviton field $h$. These are therefore interpreted as magnetic charges. The remaining $Q(K)$ are conserved only on-shell and so are interpreted as electric charges for the graviton.
In $d>4$ dimensions, a subset of the $Y_+(K)$ are co-exact and so the corresponding charges $Q(K)$ vanish identically. The remaining Penrose charges generate non-trivial 1-form symmetries of the graviton. The non-trivial $Q(K)$ are only conserved  on-shell and so are electric charges of the graviton.
There are also $(d-3)$-form symmetries which stem from the conservation of the Hodge duals of the $Y_+(K)$. In $d=4$, these `dual' charges are not independent of the Penrose charges $Q(K)$ \cite{Benedetti2023GeneralizedGravitons, Hull:2024xgo}. However, in $d>4$ dimensions, the $(d-3)$-form symmetries are related to identically conserved currents and are, therefore, magnetic charges.

In order to gauge the electric 1-form symmetries, we will show that the $Y_+(K)$ (for the relevant CKY 2-form $K$) can be interpreted as the Noether currents for transformations of the graviton field of the form
\begin{equation}\label{eq:dh=lambda_k_intro}
    \delta h_{\mu\nu} = \lambda_{(\mu} k_{\nu)}
\end{equation}
where $k$ is a Minkowski space Killing vector and $\lambda$ is a closed 1-form parameter. The relation of this transformation to the currents $Y_+(K)$ relies on several detailed properties of the CKY tensors of Minkowski space, the most important of which is that any Minkowski space Killing vector can be written as the divergence of a CKY 2-form. Given this relation, we show how to consistently couple a 2-form background gauge field $B$ to gauge this symmetry.
Clearly, it is necessary for this construction that the background spacetime admits both Killing vectors and CKY tensors.
The degree to which this structure can be generalised to linearisations of Einstein gravity around other (non-flat) background spacetimes, and to the full non-linear theory, is an important question which we will return to in future work. 

The Minkowski space Killing vectors have components
\begin{equation}
    k^\mu = V^\mu + W^{\mu\nu} x_\nu
\end{equation}
where $V$ is a constant vector, $W$ is a constant anti-symmetric rank-2 tensor, and $x^\mu$ are the standard Minkowski space coordinates. Those Killing vectors parameterised by $V$ correspond to translations, while those parameterised by $W$ correspond to Lorentz rotations. 
We find that when $k$ is a translational Killing vector, the 1-form symmetry \eqref{eq:dh=lambda_k_intro} can be gauged in the conventional manner by coupling the Noether current to a 2-form gauge field and then seeking higher order terms to achieve a gauge-invariant action.
When $k$ is a (Lorentz) rotational Killing vector, however, this  method is insufficient and it is necessary to write the Killing vector as the divergence of a CKY 2-form. We present a systematic method to gauge the symmetry in this case.

The gauging of these symmetries is related to the those discussed in \cite{Hull:2024bcl} and \cite{Hinterbichler2023GravitySymmetries}. In the former, general shifts of the graviton $\delta h_{\mu\nu} = \alpha_{\mu\nu}$ were studied. These are global symmetries of the graviton theory when $\alpha$ satisfies a certain flatness condition. A method to gauge these symmetries was found by coupling a background gauge field in the [2,1] representation of $GL(d,\mathbb{R})$.\footnote{We will refer to an irreducible representation of $GL(d,\mathbb{R})$ labelled by a Young tableau with columns of lengths $p_1$, $p_2$, $\dots$, $p_r$ as the $[p_1,p_2,\dots,p_r]$ representation.} In \cite{Hinterbichler2023GravitySymmetries}, however, a subset of these symmetries of the form $\delta h_{\mu\nu} = \partial^\rho \Lambda_{\rho\mu|\nu}$ was studied and gauged by coupling a background gauge field in the [2,2] representation. The present work instead uses a background 2-form gauge field, as is standard for 1-form symmetries.

The magnetic $(d-3)$-form symmetries can be gauged by the introduction of a background $(d-2)$-form gauge field $\tilde{B}$.
There is a dual formulation of the graviton theory which is written in terms of a gauge field in the $[d-3,1]$ representation of $GL(d,\mathbb{R})$ \cite{Hull2000}. By studying the higher-form symmetries in this dual formulation, the duality between the 1-form and $(d-3)$-form symmetries of the graviton can be made precise. We will return to this in upcoming work.

While several aspects of the higher-form symmetries of the graviton are similar to those of $p$-form gauge theories, there are salient differences. Firstly, not all pairs of 1-form and $(d-3)$-form symmetries have a mixed 't Hooft anomaly. We explicitly show that the pairs of symmetries which do have anomalies are those which were identified to be dual in \cite{Hull:2024xgo}, and are also the same pairs which were found to have non-trivial commutation relations in \cite{Benedetti2023GeneralizedGravitons}.
Another notable difference with $p$-form gauge theory is that the higher-form symmetries of the graviton (particularly the magnetic symmetries) are markedly different in $d=4$ dimensions as compared to $d>4$.
We treat these two cases in detail and find qualitatively similar results for both, however the details of the computations are different. This fact has also played an important role in the analyses of \cite{Hinterbichler2023GravitySymmetries} and \cite{Benedetti2023GeneralizedGravitons}.

The remainder of this paper is set out as follows. In section~\ref{sec:particular_shifts} the graviton theory is introduced and the global shift symmetries \eqref{eq:dh=lambda_k_intro} are analysed. The improved Penrose 2-forms are interpreted as the Noether currents associated with these symmetries and their construction is reviewed in section~\ref{sec:Penrose_review}. Particular attention is given to the properties of the CKY tensors of Minkowski space, and the set of 1-form symmetries generated by the Penrose charges. The gauging of the shift symmetries is considered in sections~\ref{sec:gauging_Chris} and \ref{sec:gauging_Max}. 
The former treats the simpler case where the shift symmetry involves a constant Killing vector, while the latter deals with the general case.
In section~\ref{sec:gauging_multiple_symms}, we consider the simultaneous gauging of multiple shift symmetries and show that there are no mixed 't Hooft anomalies amongst them. The dual $(d-3)$-form symmetries of the graviton theory and their gauging are studied in section~\ref{sec:dual_symms}. In section~\ref{sec:anomalies_with_duals}, it is shown that only certain pairs of 1-form and $(d-3)$-form symmetries can be gauged simultaneously. In the cases where this is not possible, there are mixed 't Hooft anomalies between the symmetries which are characterised by descent equations for the background fields corresponding to these generalised symmetries. The relations of these symmetries with other shift symmetries of the graviton are discussed in section~\ref{sec:comparison_with_gapless} and, finally, in section~\ref{sec:conclusion} the results are summarised and future directions outlined.
\section{A specific shift symmetry of the graviton}
\label{sec:particular_shifts}

We study a free spin-2 field $h_{\mu\nu}$ on a flat $d$-dimensional Minkowski space background $\M=\mathbb{R}^{1,d-1}$, possibly with some points or regions removed. Throughout we use the Minkowski metric $\eta_{\mu\nu} = \text{diag}(-1,1,\dots,1)$ to raise and lower indices. The graviton theory is the massless Fierz-Pauli theory \cite{Fierz:1939ix}. 
This has a gauge symmetry under linearised diffeomorphisms
\begin{equation}\label{eq:linear_diffeo}
    \delta h_{\mu\nu} = 2\partial_{(\mu} \xi_{\nu)}
\end{equation}
where $\xi$ is an arbitrary 1-form. The linearised Riemann tensor
\begin{equation}\label{eq:linear_Riemann}
    R\indices{_{\mu\nu}^{\alpha\beta}} = -2\partial_{[\mu}\partial^{[\alpha} h\indices{_{\nu]}^{\beta]}}
\end{equation}
is gauge-invariant under \eqref{eq:linear_diffeo}. The linearised Einstein tensor is defined by
\begin{equation}
    G\indices{^\mu_\nu} = -\frac{3}{2} \delta^{\mu\rho\alpha}_{\nu\sigma\beta} R\indices{_{\rho\alpha}^{\sigma\beta}}
\end{equation}
where $\delta_{\mu\rho\alpha}^{\nu\sigma\beta} = \delta^{[\nu}_\mu \delta^\sigma_\rho \delta^{\beta]}_\alpha$. The action can be written as
\begin{equation}\label{eq:graviton_action}
    S_0 = \frac{1}{2} \int\dd[d]{x} h_{\mu\nu}G^{\mu\nu}
\end{equation}
which gives rise to the field equations
\begin{equation}\label{eq:G=0}
    G_{\mu\nu} = 0
\end{equation}

In \cite{Hull:2024bcl}, the shift symmetry of the graviton theory \eqref{eq:graviton_action} under
\begin{equation}\label{eq:h_shift_general}
    \delta h_{\mu\nu} = \alpha_{\mu\nu}
\end{equation}
was studied, where $\alpha_{\mu\nu}$ is a symmetric tensor. When $\alpha$ satisfies a partial flatness condition
\begin{equation}\label{eq:G(alpha)=0}
    G(\alpha)\indices{^\mu_\nu} = 3 \delta^{\mu\rho\alpha}_{\nu\sigma\beta} \partial_\rho \partial^\sigma \alpha\indices{_\alpha^\beta} = 0
\end{equation}
and is not of the form $\alpha_{\mu\nu} = 2\partial_{(\mu} \xi_{\nu)}$, the shift \eqref{eq:h_shift_general} is a global symmetry. In \cite{Hull:2024bcl}, a method to gauge this symmetry using a gauge field in the [2,1] representation of $GL(d,\mathbb{R})$ was presented. Here, we focus on a sub-group of these symmetries, where the transformation takes the specific form
\begin{equation}\label{eq:dh=lambda_k}
    \delta h_{\mu\nu} = \lambda_{(\mu} k_{\nu)}
\end{equation}
where $k$ is a Minkowski space Killing vector. This is a global symmetry of the graviton theory \eqref{eq:graviton_action} if $\lambda$ is closed. In fact, $\lambda$ need only satisfy the weaker condition $G(\alpha)=0$, where $\alpha_{\mu\nu} = \lambda_{(\mu}k_{\nu)}$. For simplicity, we will use the stronger constraint $\dd{\lambda}=0$ here. If $\lambda$ is exact, \eqref{eq:dh=lambda_k} reduces to a gauge symmetry \eqref{eq:linear_diffeo}. Therefore, the non-trivial global symmetries of the form \eqref{eq:dh=lambda_k} are labelled by non-trivial de Rham cohomology classes $H^1(\M)$.

Under a transformation \eqref{eq:dh=lambda_k}, the graviton action \eqref{eq:graviton_action} varies by
\begin{equation}
    \delta S_0 = \int \dd[d]{x} \lambda_\mu G^{\mu\nu} k_\nu
\end{equation}
Treating the 1-form $\lambda$ as a transformation parameter, the Noether current associated with the shift \eqref{eq:dh=lambda_k} for a given $k$ is a 2-form $J(k)$ satisfying
\begin{equation}\label{eq:divJ}
    \partial^\nu J(k)_{\mu\nu} = G_{\mu\nu} k^\nu
\end{equation}
off-shell. Clearly then, $J(k)$ is conserved on-shell, when $G_{\mu\nu}=0$, and can be used to define a charge
\begin{equation}
    Q(k) = \int_{\Sigma_{d-2}} \star J(k)
\end{equation}
where $\Sigma_{d-2}$ is a codimension-2 cycle. This charge then generates a 1-form symmetry.
In the following section, we discuss a set of 1-form symmetries which have been discussed in the literature and show that these provide a natural candidate for the 2-form $J(k)$.

\section{Review of 1-form symmetries of the graviton theory}
\label{sec:Penrose_review}

We now give a brief review of a set of 
conserved 2-form currents
of the linearised graviton theory, corresponding to 1-form symmetries. A more detailed discussion can be found in \cite{Hull:2024xgo}.
The Penrose 2-form $Y(K)$ is defined as
\begin{equation}\label{eq:Penrose2-form}
    Y(K)_{\mu\nu} = R_{\mu\nu\rho\sigma} K^{\rho\sigma}
\end{equation}
where $K$ satisfies
\begin{equation}\label{eq:CKY_equation}
    \partial_\rho K_{\mu\nu} = \tilde{K}_{\rho\mu\nu} + 2 \eta_{\rho[\mu} \hat{K}_{\nu]}
\end{equation}
and
\begin{equation}
    \tilde{K}_{\rho\mu\nu} = \partial_{[\rho} K_{\mu\nu]}\qc \hat{K}_\mu = \frac{1}{d-1} \partial^\nu K_{\nu\mu}
\end{equation}
That is, $K$ is a conformal Killing-Yano (CKY) 2-form \cite{Tachibana1969OnSpace,Kashiwada1968}.
On $d$-dimensional Minkowski space, the general solution to eq.~\eqref{eq:CKY_equation} is \cite{penrose_rindler_1986,Howe2018SCKYT,Hinterbichler2023GravitySymmetries}
\begin{equation}\label{eq:CKY_solution}
    K_{\mu\nu} = \A_{\mu\nu} + \B_{[\mu} x_{\nu]} + \C_{\mu\nu\rho} x^\rho + 2x_{[\mu}\D_{\nu]\rho}x^\rho + \frac{1}{2} \D_{\mu\nu} x_\rho x^\rho
\end{equation}
where $\A$, $\B$, $\C$ and $\D$ are constant anti-symmetric tensors and $x^\mu$ are the coordinates of the background Minkowski spacetime. Therefore, the number of independent CKY 2-forms in $d$ dimensions is $d(d+1)(d+2)/6$.

The properties of each type of CKY tensor given in \eqref{eq:CKY_solution} will be important in later sections.
The $\A$-type solutions are constant 2-forms, so are both closed and co-closed (so $\hat{K}=0$ and $\tilde{K}=0$). The $\B$-type solutions are closed ($\tilde{K}=0$) but not co-closed. The $\C$-type solutions are co-closed ($\hat{K}=0$) but not closed. Finally, the $\D$-type solutions are neither closed nor co-closed.
The CKY tensors which are co-closed are often referred to as Killing-Yano tensors, and elsewhere as Yano tensors \cite{Yano1952}. For clarity, in this work we will refer to each CKY tensor as $\A$-type, $\B$-type, $\C$-type or $\D$-type.

It is a property of Minkowski space CKY 2-forms that their divergence $\hat{K}$ is a Killing vector and that all Minkowski space Killing vectors can be written in this way \cite{Hull:2024xgo}. Explicitly, \eqref{eq:CKY_solution} gives
\begin{equation}\label{eq:Khat_solution}
    \hat{K}_\mu = -\frac{1}{2} \B_\mu + \D_{\mu\nu} x^\nu
\end{equation}
The $\B$-type CKY tensors therefore give rise to constant `translational' Killing vectors, whereas the $\D$-type CKY tensors give rise to non-constant `rotational' ones.
The $\A$- and $\C$-type CKY tensors are co-closed and so do not contribute to $\hat{K}$.
There is a unique $\B$- or $\D$-type CKY 2-form related to each Minkowski space Killing vector by \eqref{eq:Khat_solution}.

The Penrose 2-form is conserved on-shell:
\begin{equation}
    \partial^\nu Y(K)_{\mu\nu} \doteq 0
\end{equation}
where $\doteq$ denotes an on-shell equality. This follows from \eqref{eq:CKY_equation} and the contracted Bianchi identity $\partial^\mu R_{\mu\nu\rho\sigma} = 2 \partial_{[\rho} R_{\sigma]\nu}$ since the Ricci tensor vanishes on the equations of motion \eqref{eq:G=0}. 

Furthermore, when $K$ is closed (either $\A$- or $\B$-type), $Y(K)$ is also closed off-shell:
\begin{equation}
    \partial_{[\mu} Y(K)_{\nu\rho]} = \partial_{[\mu} R_{\nu\rho]\alpha\beta}K^{\alpha\beta} + R_{\alpha\beta[\nu\rho}\partial_{\mu]}K^{\alpha\beta} = - 2 R_{\beta[\mu\nu\rho]} \hat{K}^\beta = 0
\end{equation}
where we have used \eqref{eq:CKY_equation} with $\tilde{K}=0$ as well as the Bianchi identities
\begin{equation}\label{eq:differential_Bianchi}
    \partial_{[\mu}R_{\nu\rho]\alpha\beta} = 0
\end{equation}
and
\begin{equation}\label{eq:algebraic_Bianchi}
    R_{\beta[\mu\nu\rho]}=0
\end{equation}
The improved Penrose current is the 2-form\footnote{If $K$ is taken to be an $\A$- or $\C$-type CKY tensor (i.e. a Killing-Yano tensor) then $Y_+(K)$ is proportional to the linearisation of the current considered in \cite{Kastor2004ConservedTensors} for n=2.  }
\begin{equation}\label{eq:improved_Penrose}
    Y_+(K)_{\mu\nu} = R_{\mu\nu\rho\sigma}K^{\rho\sigma} + 4R\indices{^\rho_{[\mu}} K_{\nu]\rho} + RK_{\mu\nu} = 6 \delta_{\mu\nu\alpha\beta}^{\rho\sigma\gamma\delta} K_{\rho\sigma} R\indices{^{\alpha\beta}_{\gamma\delta}}
\end{equation}
where $\delta_{\mu\nu\alpha\beta}^{\rho\sigma\gamma\delta} = \delta^{[\mu}_\rho \delta^\nu_\sigma \delta^\alpha_\gamma \delta^{\beta]}_\delta$. We note that on-shell $Y(K) \doteq Y_+(K)$.
The improved Penrose 2-form satisfies
\begin{equation}\label{eq:divPenrose}
    \partial^\nu Y_+(K)_{\mu\nu} = 2(d-3) G_{\mu\nu}\hat{K}^\nu
\end{equation}
where we have used the Bianchi identities \eqref{eq:differential_Bianchi} and \eqref{eq:algebraic_Bianchi}.
So $Y_+(K)$ is also conserved on-shell, as was guaranteed.

Returning to the particular graviton shifts \eqref{eq:dh=lambda_k}, we saw that the associated Noether current is a 2-form satisfying \eqref{eq:divJ}. Indeed, from \eqref{eq:divPenrose}, the improved Penrose 2-form $Y_+(K)$ satisfies this relation provided that we set
\begin{equation}\label{eq:k=Khat}
    k = 2(d-3) \hat{K}
\end{equation}
As remarked above, all Minkowski space Killing vectors can be written uniquely as the divergence $\hat{K}$ of a $\B$- or $\D$-type CKY 2-form $K$.
Therefore, we can interpret the improved Penrose 2-form \eqref{eq:improved_Penrose} as the Noether current associated with the global symmetry \eqref{eq:dh=lambda_k}, where $\dd\lambda=0$. 
We will restrict $K$ to be either $\B$- or $\D$-type, since the $\A$- and $\C$-type CKY 2-forms have $\hat{K}=0$ and so 
would result in a transformation $\delta h_{\mu\nu} = 0$.

The Penrose charges $Q(K)$ are defined as integrals of the improved Penrose current
\begin{equation}\label{eq:PenroseCharge}
    Q(K) = \int_{\Sigma_{d-2}} \star Y_+(K)
\end{equation}
where $\Sigma_{d-2}$ is a $(d-2)$-cycle. The conservation of $Y_+(K)$ implies that $Q(K)$ is a topological operator (i.e. it does not vary under small deformations of $\Sigma_{d-2}$ which do not cross any other insertions) and so generates a 1-form symmetry for each CKY tensor $K$.

In $d=4$, the Penrose charges associated with all 20 CKY 2-forms are non-vanishing in general, and generate a group $\mathbb{R}^{20}$ of 1-form symmetries. In $d>4$, however, the Penrose charges constructed from $\A$- and $\C$-type CKY 2-forms vanish \cite{Benedetti2023GeneralizedGravitons, Hull:2024xgo}, and the remaining $\B$- and $\D$-type Penrose charges generate a group $\mathbb{R}^{d(d+1)/2}$ of 1-form symmetries. 
From \eqref{eq:divPenrose}, these charges are only conserved on-shell and so should be interpreted as electric charges.

In $d=4$, the $\A$- and $\C$-type Penrose charges do not vanish but, since $\hat{K}=0$ for these CKY tensors, they cannot be interpreted as the Noether charges for any transformation of the field $h_{\mu\nu}$. Equivalently, the improved Penrose 2-form $Y_+(K)$ is conserved identically (not only on-shell) when $\hat{K}=0$, and so the $\A$- and $\C$-type Penrose charges should be understood as generating magnetic symmetries in four dimensions.\footnote{This is analogous to the magnetic $U(1)$ 1-form symmetry of four-dimensional Maxwell theory, whose Noether current is proportional to $\star F$. This is identically conserved by the Bianchi identity and cannot be understood as the Noether current associated with any continuous transformation of the field $A_\mu$.} We will return to these magnetic symmetries  in section \ref{sec:dual_symms_4d}.

A complementary perspective on the Penrose charges was given in \cite{Hull:2024xgo} where they were found to be related to the linearisation of the ADM charges \cite{ADM, Abbott1982StabilityConstant}, as well as a set of magnetic charges. These magnetic charges stem from identically conserved 2-form currents in the graviton theory, and can be related to electric charges of the dual graviton \cite{Hull:2023iny}. In particular, the viewpoint advocated in \cite{Hull:2024xgo} was that the improved Penrose 2-form can be seen as a gauge-invariant improvement of the standard 2-form current of \cite{Abbott1982StabilityConstant} which leads to the ADM charges.

\section{Gauging 1-form symmetries with constant \texorpdfstring{$k$}{k}}
\label{sec:gauging_Chris}

From the previous section, the Penrose charges \eqref{eq:PenroseCharge} can be seen as  Noether charges associated with the shift symmetries \eqref{eq:dh=lambda_k} where $\lambda$ is a closed (and non-exact) 1-form and $k$ is a Killing vector. These are global symmetries as there is a constraint $\dd\lambda=0$ on the transformation parameter. In order to diagnose a 't Hooft anomaly amongst these symmetries, it is interesting to study their gauging.

Let us begin with the  case in which the Killing vector $k$ is a constant vector. 
In this case, we will see that it is possible to couple a background 2-form gauge field without using the relation of Killing vectors to CKY 2-forms that we found in the previous section. This is not the case when $k$ is a non-constant Killing vector, as we will see in section~\ref{sec:gauging_Max}.

\subsection{Gauge-invariant field equation}

It is simple to write down a gauge-invariant field equation as follows. Firstly, the linearised Riemann tensor has components given in \eqref{eq:linear_Riemann} and so under the variation \eqref{eq:dh=lambda_k}, it varies as
\begin{equation}
    \delta R_{\mu\nu\rho\sigma} = k_{[\rho} \partial_{\sigma]} \partial_{[\mu} \lambda_{\nu]} + k_{[\mu} \partial_{\nu]} \partial_{[\rho} \lambda_{\sigma]}
\end{equation}
when $k$ is a constant. The relevant background gauge field to introduce is a 2-form gauge field $B_{\mu\nu}$ transforming as
\begin{equation}\label{eq:B_transf}
    B_{\mu\nu} \to B_{\mu\nu} + \partial_{[\mu} \lambda_{\nu]}
\end{equation}
We define the modified Riemann tensor
\begin{equation}\label{eq:modified_Riemann_constant_k}
    \mathcal{R}_{\mu\nu\rho\sigma} = R_{\mu\nu\rho\sigma} - k_{[\rho} \partial_{\sigma]} B_{\mu\nu} - k_{[\mu} \partial_{\nu]} B_{\rho\sigma}
\end{equation}
which is invariant under combined transformations \eqref{eq:dh=lambda_k} and \eqref{eq:B_transf}, i.e.
\begin{equation}\label{eq:electric_transf}
    \delta h_{\mu\nu} = \lambda_{(\mu} k_{\nu)} \qc \delta B_{\mu\nu} =  \partial_{[\mu} \lambda_{\nu]}
\end{equation}
Therefore, a gauge-invariant field equation is 
\begin{equation}\label{eq:gauge_invt_field_eqn}
    \mathcal{G}_{\mu\nu} = G_{\mu\nu} + 3 \eta_{\rho\alpha(\mu|\nu)\sigma\beta} k^\rho \partial^\alpha B^{\sigma\beta} = 0
\end{equation}
where 
\begin{equation}\label{eq:modified_Einstein}
    \mathcal{G}_{\mu\nu} = -\frac{3}{2} \eta_{\mu\rho\alpha|\nu\sigma\beta} \mathcal{R}^{\rho\alpha\sigma\beta}
\end{equation}
is the Einstein tensor constructed from $\mathcal{R}$ instead of $R$, and
\begin{equation}
    \eta_{\mu\nu\rho|\alpha\beta\gamma} = \eta_{\mu\kappa} \eta_{\nu\lambda} \eta_{\rho\sigma} \delta^{\kappa\lambda\sigma}_{\alpha\beta\gamma}
\end{equation}
This tensor satisfies the important properties
\begin{equation}
    \eta_{\mu\nu\rho|\alpha\beta\gamma} = \eta_{\alpha\beta\gamma|\mu\nu\rho}\qc \eta_{[\mu\nu\rho|\alpha]\beta\gamma} = 0
\end{equation}
In other words, it transforms in the [3,3] representation of $GL(d,\mathbb{R})$.

We note that $\mathcal{R}$ does not satisfy the analogues of the Bianchi identities, that is,
\begin{align}
    \mathcal{R}_{[\mu\nu\alpha]\beta} &= -2 k_{[\mu} \partial_\nu B_{\alpha\beta]} \label{eq:gauged_non_bianchi1} \\
    \partial_{[\sigma} \mathcal{R}_{\mu\nu]\alpha\beta} &= -k_{[\alpha} \partial_{\beta]} \partial_{[\sigma} B_{\mu\nu]} \label{eq:gauged_non_bianchi2}
\end{align}
In the case where the background field $B$ is flat, i.e. $\dd{B}=0$, then the right-hand side of both \eqref{eq:gauged_non_bianchi1} and \eqref{eq:gauged_non_bianchi2} vanish and the familiar Bianchi identities for the Riemann tensor in the ungauged theory hold true also in the gauged theory for the modified Riemann tensor \eqref{eq:modified_Riemann_constant_k}.

\subsection{Gauge-invariant action}

It is also possible to introduce couplings involving $B$ into the graviton action \eqref{eq:graviton_action} to make it gauge-invariant under \eqref{eq:electric_transf} (with $k$ a constant vector).
Integrating the graviton action \eqref{eq:graviton_action} by parts, it can be written
\begin{equation}
    S_0 = -\frac{3}{2} \int \dd[d]{x} \delta_{\mu\rho\alpha}^{\nu\sigma\beta} \partial^\rho h\indices{^\mu_\nu} \partial_\sigma h\indices{^\alpha_\beta}
\end{equation}
Under a transformation \eqref{eq:dh=lambda_k}, it varies by
\begin{equation}\label{eq:dSFP}
    \delta S_0 = -3\int \dd[d]{x} \delta^{\nu\sigma\beta}_{\mu\rho\alpha} \partial^\rho h\indices{^\mu_\nu} k^\alpha \partial_{[\sigma} \lambda_{\beta]}
\end{equation}
to first order in $\lambda$, where we have used that $k$ is a constant. Therefore, we introduce a coupling between $h$ and $B$
\begin{equation}\label{eq:first_order_coupling_constant_k}
    S_0' = S_0 + 3\int\dd[d]{x} \delta^{\nu\sigma\beta}_{\mu\rho\alpha} \partial^\rho h\indices{^\mu_\nu} k^\alpha B_{\sigma\beta}
\end{equation}
such that the variation \eqref{eq:dSFP} is cancelled by the variation of $B$ in this new term. However, the new term also varies under \eqref{eq:dh=lambda_k}.
This variation can be cancelled by the addition of a quadratic counter term depending only on the background field $B$. The full gauged action is
\begin{equation}\label{eq:S'_constant_k}
    S' = \int\dd[d]{x} \left( \frac{1}{2} h_{\mu\nu} G^{\mu\nu} + 3 \delta^{\nu\sigma\beta}_{\mu\rho\alpha} \partial^\rho h\indices{^\mu_\nu} k^\alpha B_{\sigma\beta} - \frac{3}{4} \delta^{\nu\sigma\beta}_{\mu\rho\alpha} B^{\rho\mu} k_\nu k^\alpha B_{\sigma\beta} \right)
\end{equation}
While the variations above were calculated only to first order in $\lambda$, it is simple to verify that this action is exactly gauge-invariant under the combined transformations \eqref{eq:dh=lambda_k} and \eqref{eq:B_transf}.
The field equation for $h$ which follows from this action is \eqref{eq:gauge_invt_field_eqn}. For this simple case where $k$ is a constant Killing vector, \eqref{eq:S'_constant_k} gives a way to couple a background field for the symmetry \eqref{eq:dh=lambda_k} without using the  extra structure of the CKY tensors and Penrose 2-form. We will see in the next section that for general $k$, this simple method is insufficient and we must make use of the extra structure provided by the CKY tensors.
\section{The general case: gauging 1-form symmetries}
\label{sec:gauging_Max}

In the previous section, we saw that the shift symmetry \eqref{eq:dh=lambda_k} with $k$ constant can be gauged by the introduction of a 2-form background gauge field $B$ simply by analysing the variation of the graviton action under the transformation and introducing counterterms. In that case, the variations were sufficiently simple that they could be written in terms of $\dd\lambda$ and then gauged immediately by introducing the obvious couplings of $B$.

We now treat the general case where $k$ is an arbitrary Minkowski space Killing vector, and so can be at most linear in the coordinates $x^\mu$. We will make frequent use of the fact that the Killing vectors satisfy $\partial_{(\mu} k_{\nu)} = 0$ and $\partial_\mu \partial_\nu k_\alpha = 0$. 

\subsection{Gauge-invariant field equation}

Identifying a gauge-invariant field equation is possible by again examining the variation of the Riemann tensor, which can be written
\begin{align}
\begin{split}\label{eq:Riemann_variation}
    \delta R_{\mu\nu\alpha\beta} &= k_{[\alpha} \partial_{\beta]} \partial_{[\mu} \lambda_{\nu]} + k_{[\mu} \partial_{\nu]} \partial_{[\alpha} \lambda_{\beta]} - \partial_{[\mu} \lambda_{\nu]} \partial_\alpha k_\beta - \partial_{[\alpha} \lambda_{\beta]} \partial_\mu k_\nu \\
    & \quad - \frac{1}{2} \partial_{[\alpha} \lambda_{\nu]} \partial_\mu k_\beta - \frac{1}{2} \partial_{[\mu} \lambda_{\beta]} \partial_\alpha k_\nu + \frac{1}{2} \partial_{[\beta} \lambda_{\nu]} \partial_\mu k_\alpha + \frac{1}{2} \partial_{[\mu} \lambda_{\alpha]} \partial_\beta k_\nu
\end{split}
\end{align}
Since this depends only on $\dd\lambda$, we can define a modified Riemann tensor
\begin{equation}\label{eq:modified_Riemann}
    \mathcal{R}'_{\mu\nu\alpha\beta} = R_{\mu\nu\alpha\beta}  - k_{[\alpha} \partial_{\beta]} B_{\mu\nu} - k_{[\mu} \partial_{\nu]} B_{\alpha\beta} + B_{\mu\nu} \partial_\alpha k_\beta + B_{\alpha\beta} \partial_\mu k_\nu - \left( B_{\alpha[\mu} \partial_{\nu]} k_\beta - \alpha\leftrightarrow\beta \right)
\end{equation}
which is gauge-invariant. 
Indeed, if $k$ is a constant vector then this reduces to the modified Riemann tensor $\mathcal{R}$ found previously in \eqref{eq:modified_Riemann_constant_k}. 
A gauge-invariant modification of the Einstein tensor is
\begin{equation}\label{eq:modified_Einstein2}
    \mathcal{G}'_{\mu\nu} = -\frac{3}{2} \eta_{\mu\rho\alpha|\nu\sigma\beta} \mathcal{R'}^{\rho\alpha\sigma\beta}
\end{equation}
which evaluates to
\begin{equation}\label{eq:gauge_invt_field_eqn2}
    \mathcal{G}'_{\mu\nu} = G_{\mu\nu} - 3\eta_{\rho\alpha(\mu|\nu)\sigma\beta} \left( -k^\sigma \partial^\beta B^{\rho\alpha} + \frac{3}{2} B^{\rho\alpha}\partial^\sigma k^\beta \right)
\end{equation}

As in the case considered in the previous section, $\mathcal{R}'$ does not satisfy the analogs of the Bianchi identities:
\begin{align}
    \mathcal{R}'_{[\mu\nu\alpha]\beta} &= -2k_{[\mu} \partial_\nu B_{\alpha\beta]} \label{eq:gauged_non_bianchi3} \\
    \partial_{[\sigma} \mathcal{R}'_{\mu\nu]\alpha\beta} &= -k_{[\alpha} \partial_{\beta]} \partial_{[\sigma} B_{\mu\nu]} + 2\partial_{[\sigma} B_{\mu\nu]} \partial_\alpha k_\beta - 2 \left( \partial_{[\sigma} B_{\mu\nu} \partial_{\alpha]} k_\beta - \alpha \leftrightarrow \beta \right) \label{eq:gauged_non_bianchi4}
\end{align}
We note that the Bianchi identities are again satisfied when the background is flat, $\dd{B}=0$.

\subsection{Gauge-invariant action}

It is more challenging to construct a gauge-invariant action. Recall our earlier observation that the improved Penrose current $Y_+(K)$, defined in \eqref{eq:improved_Penrose}, can be taken as the Noether current for the shift symmetry \eqref{eq:dh=lambda_k}, where $K$ and $k$ are related by \eqref{eq:k=Khat}. We begin by adding a standard coupling of the Noether current to the background field $B$ to the graviton action \eqref{eq:graviton_action},
\begin{equation}\label{eq:first_order_gauge}
    S_0'' = S_0 - \int\dd[d]{x} B_{\mu\nu} Y_+(K)^{\mu\nu}
\end{equation}
such that the variation of $S_0$ is cancelled by the variation of $B$ in the final term. However, since $Y_+(K)$ is itself a function of $h$, it transforms under the shift \eqref{eq:dh=lambda_k}. Our task is now to identify a counterterm quadratic in the background field $B$ whose variation precisely cancels the variation of the final term in \eqref{eq:first_order_gauge}. Such a counterterm can be identified and a systematic approach to determining it is given in Appendix~\ref{app:counterterms}. The final gauged action can be written
\begin{equation}\label{eq:gauged_action}
    S'' = S_0 + s(k,B)
\end{equation}
where 
\begin{equation}\label{eq:s(k)_def}
    s(k,B) = \int\dd[d]{x} \left( - B_{\mu\nu} Y_+(K)^{\mu\nu} + f(K,B) \right)
\end{equation}
with $f(K,B)$ a quadratic function of the 2-form background gauge field $B$ given in \eqref{eq:f(B)}. The action $S''$ is now invariant under the combined transformations \eqref{eq:electric_transf} and \eqref{eq:B_transf} for all Killing vectors $k$ and an arbitrary 1-form $\lambda$.

The field equation for $h$ which follows from this gauged action can be written
\begin{equation}\label{eq:gauged_eom}
    \mathcal{G'}_{\mu\nu} + 6 \left( (\eta_{\nu[\mu} K_{\alpha\beta} \partial_{\gamma]} \partial^{[\gamma} B^{\alpha\beta]} - \partial_{[\mu} B_{\alpha\beta]} \tilde{K}\indices{^{\alpha\beta}_\nu}) + \mu\leftrightarrow\nu \right) + 4\eta_{\mu\nu} \partial_{[\alpha} B_{\beta\gamma]} \tilde{K}^{\alpha\beta\gamma} = 0
\end{equation}

\subsection{Comparison with the constant \texorpdfstring{$k$}{k} gauging}

In section~\ref{sec:gauging_Chris}, we found a simple way to gauge the global \eqref{eq:dh=lambda_k} symmetry when $k$ is a constant Killing vector. In the present section, we introduced a more general, yet computationally more complicated, method which allowed us to gauge this symmetry for an arbitrary Killing vector $k$. We now discuss how this more general method relates to the one in section~\ref{sec:gauging_Chris} in the case where $k$ is a constant vector.

First, let us consider the terms in the gauged actions which are linear in the background field $B$. For the simpler gauging of section~\ref{sec:gauging_Chris}, this is given by $S_0'$ in \eqref{eq:first_order_coupling_constant_k}, whereas for the more general gauging procedure of the present section, the linear coupling is given by $S_0''$ in \eqref{eq:first_order_gauge}. Both of these couplings can be written in the form
\begin{equation}\label{eq:linear_coupling}
    S_0 + \int\dd[d]{x} J_{\text{N}}^{\mu\nu} B_{\mu\nu}
\end{equation}
where $J_{\text{N}}^{\mu\nu}$ is a conserved current. Indeed, for a continuous 1-form symmetry, this is the standard coupling between the 2-form Noether current $J_{\text{N}}$ and a 2-form background gauge field.\footnote{In the familiar case of a continuous 0-form global symmetry, the Noether current $J_{\text{N}}$ is a 1-form and, if there is no 't Hooft anomaly, the symmetry can be gauged by introducing a background 1-form gauge field $A$ with a coupling $\int\dd[d]{x} J_{\text{N}}^\mu A_\mu$ to linear order in $A$. Equation \eqref{eq:linear_coupling} is the analogue of this for 1-form symmetries.} 
The difference between the couplings to the background field $B$ in $S_0'$ and $S_0''$ could be described as choosing a different Noether current $J_{\text{N}}$ in the two cases. In $S_0'$, the Noether current is
\begin{equation}
    J_{\text{N}}'^{\mu\nu} = 3 \delta^{\mu\nu\beta}_{\sigma\gamma\delta} \partial^\sigma h\indices{^\delta_\beta} k^\gamma
\end{equation}
whereas in $S_0''$, the Noether current is taken as
\begin{equation}
    J_{\text{N}}''^{\mu\nu} = -Y_+(K)^{\mu\nu}
\end{equation}
The Noether current for a given symmetry is not uniquely defined, and any co-exact 2-form can be added to $J_{\text{N}}$ without spoiling its conservation. Indeed, the two Noether currents $J_{\text{N}}'$ and $J_{\text{N}}''$ are related by such a co-exact term:
\begin{equation}\label{eq:Noether_current_relation}
    J_{\text{N}}'^{\mu\nu} = J_{\text{N}}''^{\mu\nu} - 12 \partial_\alpha \left( \delta^{\mu\nu\alpha\beta}_{\rho\sigma\gamma\delta} K^{\rho\sigma} \partial^\gamma h\indices{_\beta^\delta} \right)
\end{equation}
In deriving this result, we have used the fact that when $k$ is a constant vector, the CKY 2-form $K$ which relates to it by \eqref{eq:k=Khat} is $\B$-type and so has $\tilde{K}=0$. 

The  Noether current $J_{\text{N}}''$ (i.e. the improved Penrose 2-form)   is a gauge-invariant 2-form while
  the Noether current $J_{\text{N}}'$ is related to the ADM charges in the linearised theory \cite{Abbott1982StabilityConstant}. The relation \eqref{eq:Noether_current_relation} between these two   currents has been explored   in \cite{Hull:2024xgo} where it was shown that the Penrose charges encode both the ADM charges as well as a subset of the magnetic charges for the graviton discovered in \cite{Hull:2023iny}. These magnetic charges relate to the co-exact term in \eqref{eq:Noether_current_relation}. Furthermore, in \cite{Hull:2024mfb}, it was shown that on topologically non-trivial spacetimes there is a further set of 1-form symmetries beyond those generated by the Penrose charges, and that these relate to the remaining magnetic charges found in \cite{Hull:2023iny} by a similar relation involving the addition of co-exact terms. We will not consider these in this work.

From \eqref{eq:Noether_current_relation}, the linear couplings in $S_0'$ and $S_0''$ can be related by
\begin{equation}\label{eq:linear_coupling_relation}
    S_0' = S_0'' + 12 \int \dd[d]{x} \delta^{\mu\nu\alpha\beta}_{\rho\sigma\gamma\delta} \partial_{[\alpha}B_{\mu\nu]} K^{\rho\sigma} \partial^\gamma h\indices{_\beta^\delta}
\end{equation}
where we have integrated the co-exact term in \eqref{eq:Noether_current_relation} by parts. We see that the effect of choosing a different Noether current is to introduce a term into the action which depends on $B$ only through its curvature $\dd{B}$. Since $\dd{B}$ is invariant under the background gauge transformation \eqref{eq:B_transf}, the variation of this term under the combined transformations \eqref{eq:dh=lambda_k} and \eqref{eq:B_transf} is schematically of the form $\int \dd{B} K \partial(\lambda k)$ and must be cancelled by the addition of a counter-term quadratic in $B$. 
Since the Noether currents $J_{\text{N}}'$ and $J_{\text{N}}''$ are not invariant under the transformation \eqref{eq:dh=lambda_k}, their variations must also be cancelled by such quadratic counter-terms.
The quadratic terms required in the two methods presented in section~\ref{sec:gauging_Chris} and the present section are different. This stems from the fact that, in the latter case, these quadratic terms must also cancel the variation of the extra term in \eqref{eq:linear_coupling_relation}.
This accounts for the difference between the quadratic counter-terms in the gauged actions \eqref{eq:S'_constant_k} and \eqref{eq:gauged_action} which, by construction, cancel the variations of the Noether currents $J_{\text{N}}'$ and $J_{\text{N}}''$ respectively.
\section{Gauging multiple 1-form symmetries}
\label{sec:gauging_multiple_symms}

Equation \eqref{eq:gauged_action} gives a way to consistently couple a 2-form background gauge field $B$ to the graviton action which gauges the shift symmetry \eqref{eq:dh=lambda_k} for a given Killing vector $k$.
We now ask if multiple shift symmetries can be simultaneously gauged in this manner.
Let $k_I$ with $I=1,\dots,d(d+1)/2$ be a set of linearly independent Killing vectors of the background Minkowski space. The index $I$ then labels the independent electric 1-form symmetries \eqref{eq:dh=lambda_k}.
Consider a pair of Killing vectors $k_I$ and $k_J$ with $I\neq J$. 
We wish to consistently couple two 2-form background fields $B^I$ and $B^J$ to gauge the 1-form symmetries labelled by $k_I$ and $k_J$ respectively. 
The resulting theory should then have two gauge symmetries,\footnote{Throughout, we will use the convention that repeated $I$ indices are not summed over unless explicitly indicated.}
\begin{equation}\label{eq:kI_transf}
    \delta h_{\mu\nu} = \frac{1}{2} \left( \lambda^I_{\mu} k_{I\nu} + \lambda^I_{\nu} k_{I\mu} \right) \qc \delta B^I_{\mu\nu} = \partial_{[\mu} \lambda^I_{\nu]}
\end{equation}
and
\begin{equation}\label{eq:kJ_transf}
    \delta h_{\mu\nu} = \frac{1}{2} \left( \lambda^J_{\mu} k_{J\nu} + \lambda^J_{\nu} k_{J\mu} \right) \qc \delta B^J_{\mu\nu} = \partial_{[\mu} \lambda^J_{\nu]}
\end{equation}
with $\lambda^I$ and $\lambda^J$ unconstrained 1-form parameters.
Na\"{i}vely, the gauging would involve the addition of the same couplings found in the previous section, i.e.
\begin{align}
    S_1 &= S_0 + s(k_I,B^I) + s(k_J,B^J) \label{eq:naive_action} \\
    &= \int\dd[d]{x} \left( \frac{1}{2} h_{\mu\nu} G^{\mu\nu} - B^I_{\mu\nu} Y_+(K_I)^{\mu\nu} + f(K_I,B^I) - B^J_{\mu\nu} Y_+(K_J)^{\mu\nu} + f(K_J,B^J) \right) \nonumber
\end{align}
with $s(k,B)$ defined in \eqref{eq:s(k)_def}, and $K_I$, $K_J$ are $\B$- or $\D$-type CKY tensors related to $k_I$ and $k_J$ respectively by \eqref{eq:k=Khat}. 
However, $S_1$ is invariant under neither \eqref{eq:kI_transf} nor \eqref{eq:kJ_transf} since the $B^J_{\mu\nu} Y_+(K_J)^{\mu\nu}$ term transforms non-trivially under \eqref{eq:kI_transf} and $B^I_{\mu\nu} Y_+(K_I)^{\mu\nu}$ transforms under \eqref{eq:kJ_transf}. 
We show in Appendix~\ref{app:gauging_multiple_electric_symmetries} that it is possible to find counter-terms depending on $B^I$ and $B^J$ which give an action which is simultaneously invariant under both \eqref{eq:kI_transf} and \eqref{eq:kJ_transf}, for any choice of Killing vectors $k_I$ and $k_J$. The resulting action can be written
\begin{equation}\label{eq:S1''_final_form}
    S_1'' = S_0 + s(k_I,B^I) + s(k_J,B^J) + s(k_I,B^I,k_J,B^J)
\end{equation}
where $s(k_I,B^I,k_J,B^J)$ is defined in \eqref{eq:s(k,k')_def} and contains counter-terms depending only on the background fields $B^I$ and $B^J$. Furthermore, as shown in Appendix~\ref{app:exchange_symmetry}, $s(k_I,B^I,k_J,B^J)$ is symmetric under the exchange of $(k_I,B^I) \leftrightarrow (k_J,B^J)$. Therefore, the full action $S_1''$ is also symmetric under this exchange.

We recall from the discussion in section~\ref{sec:Penrose_review} that the 1-form symmetries generated by the Penrose charges in $d>4$ dimensions are electric symmetries and form a group $\mathbb{R}^{d(d+1)/2}$.
It follows from the discussion above that any sub-group of the electric 1-form symmetry group $\mathbb{R}^{d(d+1)/2}$ can be gauged by the introduction of the relevant counter-terms. Consider a subgroup generated by Killing vectors $k_I$ with $I \in \mathcal{I} \subseteq \{ 1,\dots,d(d+1)/2 \}$. Then an action which is gauge-invariant under \eqref{eq:kI_transf} for all $I\in\mathcal{I}$ is
\begin{equation}
    S_\mathcal{I} = S_0 + \sum_{I\in\mathcal{I}} s(k_I,B^I) + \sum_{\substack{I,J\in\mathcal{I}\\ I<J}} s(k_I,B^I,k_J,B^J)
\end{equation}
Therefore, there are no mixed 't Hooft anomalies amongst the electric 1-form symmetries.

\section{Dual symmetries and their gauging}
\label{sec:dual_symms}

We now discuss dual `magnetic' symmetries present in the graviton theory. These are $(d-3)$-form symmetries which behave differently in $d=4$ and $d>4$.

\subsection{Four dimensions}
\label{sec:dual_symms_4d}

In $d=4$, there are 20 one-form symmetries generated by the Penrose charges. Ten of these are generated by Penrose charges corresponding to CKY tensors of the $\B$ and $\D$ types. In section~\ref{sec:Penrose_review} we saw that these can be interpreted as Noether charges associated with shift symmetries $\delta h_{\mu\nu} = \lambda_{(\mu}k_{\nu)}$ where $k$ is related to the CKY tensors by \eqref{eq:k=Khat}. In analogy to the 1-form symmetry generated by $\int \star F$ in Maxwell theory, these are `electric' symmetries.

The Penrose charges corresponding to the $\A$- and $\C$-type CKY tensors, of which there are also ten, are not related to any field transformation. 
In this sense, they are interpreted as `magnetic' (or topological) symmetries, much like the 1-form symmetry generated by $\int F$ in Maxwell theory. That is, the improved Penrose currents corresponding to these CKY tensors are conserved identically, not just on solutions to the field equations.

The electric and magnetic higher-form symmetries of Maxwell theory are exchanged under duality. This is true of the $\B$- and $\C$-type Penrose charges in the graviton theory.
In the dual graviton theory, the $\C$-type shift symmetries are related to a shift of the dual graviton, whereas the $\B$-type ones are not. Interestingly, the $\A$-type symmetries cannot be related to a field transformation in either formulation of the theory and so appear as `magnetic' charges in both formulations. On the other hand, the $\D$-type symmetries are `electric' in both formulations.

In four dimensions, the Hodge duals of the improved Penrose currents $\star Y_+(K)$ are also conserved on-shell \cite{Benedetti2023GeneralizedGravitons, Hull:2024xgo}. However, due to the properties of CKY tensors under Hodge duality, the charges found by integrating them are not independent of the original Penrose charges. So the independent 1-form symmetries are the 20 Penrose charges.

Each of the magnetic symmetries can be gauged by the introduction of a standard coupling between the Penrose current and a 2-form background gauge field $B'$,
\begin{equation}\label{eq:4d_AC_coupling}
    S_2^{d=4} = \int \dd[4]{x} \left( \frac{1}{2}h_{\mu\nu}G^{\mu\nu} + B'_{\mu\nu} Y_+(K')^{\mu\nu} \right)
\end{equation}
where $K'$ is an $\A$- or $\C$-type CKY tensor.
The background field transforms as 
\begin{equation}\label{eq:B'_background_transf}
    \delta B'_{\mu\nu} = \partial_{[\mu} \lambda'_{\nu]}
\end{equation}
This is a gauge invariance of $S^{d=4}_2$ since $\partial_\mu Y_+(K')^{\mu\nu}=0$ even off-shell.

\subsection{\texorpdfstring{$d>4$}{d>4} dimensions}

In higher dimensions, the Penrose charges associated with $\A$- and $\C$-type CKY tensors vanish when defined on codimension-2 cycles (so they act as the identity on all line operators they link with) \cite{Benedetti2023GeneralizedGravitons, Hull:2024xgo}, leaving only the $\B$- and $\D$-type Penrose charges generating a group $\mathbb{R}^{d(d+1)/2}$ of non-trivial 1-form symmetries. There are, however, dual $(d-2)$-form currents which are the Hodge duals $\star Y(K'')$ of the Penrose 2-form in \eqref{eq:Penrose2-form}.\footnote{Note that this is the Hodge dual of the Penrose 2-form, \emph{not} the improved Penrose 2-form.} These are identically conserved when $K''$ is an $\A$- or $\B$-type CKY tensor. (This is not to be confused with $K'$ in the previous subsection, which is an $\A$- or $\C$-type CKY tensor.) These conserved $(d-2)$-form currents can be integrated to give the dual Penrose charges
\begin{equation}\label{eq:dual_Penrose_charges}
    \tilde{Q}(K'') = \int_{\Sigma_2} Y(K'')
\end{equation}
where $\Sigma_2$ is a 2-cycle, which then generate a group $\mathbb{R}^{d(d+1)/2}$ of $(d-3)$-form symmetries. 
These magnetic symmetries can be gauged by adding a coupling to a background $(d-2)$-form gauge field $\tilde{B}$
\begin{equation}\label{eq:AC_coupling_d>4}
\begin{split}
    S_2^{d>4} &= \int \dd[d]{x} \left( \frac{1}{2} h_{\mu\nu} G^{\mu\nu} + \frac{1}{(d-2)!} \tilde{B}_{\mu_1\dots\mu_{d-2}} (\star Y(K''))^{\mu_1\dots\mu_{d-2}} \right) \\
    &= \frac{1}{2} \int \dd[d]{x} h_{\mu\nu} G^{\mu\nu} - \int \tilde{B} \wedge Y(K'')
\end{split}
\end{equation}
where $K''$ is an $\A$- or $\B$-type CKY tensor and the background field transforms as
\begin{equation}\label{eq:Btilde_background_transf}
    \delta \tilde{B}_{\mu_1\dots\mu_{d-2}} = \partial_{[\mu_1} \tilde{\lambda}_{\mu_2\dots\mu_{d-2}]}
\end{equation}

\subsection{Gauging multiple \texorpdfstring{$(d-3)$}{(d-3)}-form symmetries}

In section \ref{sec:gauging_multiple_symms}, we focused on the electric 1-form symmetries related to the $\B$- and $\D$-type CKY tensors, which correspond to shift symmetries \eqref{eq:dh=lambda_k}, and showed that any pair of them can be simultaneously gauged.
It is straightforward to see that the same is true of the magnetic symmetries introduced above. 

Let us first consider $d=4$, where there are 10 magnetic 1-form symmetries corresponding to the $\A$- and $\C$-type CKY tensors. Each one of these can be gauged as in \eqref{eq:4d_AC_coupling}. Introducing two couplings of this sort immediately gives an action which is simultaneously invariant under both background transformations. Explicitly, consider the action
\begin{equation}\label{eq:4d_two_magnetic_gauged}
    S_3^{d=4} = \int \dd[4]{x} \left( \frac{1}{2}h_{\mu\nu}G^{\mu\nu} +  B'^I_{\mu\nu} Y_+(K'_I)^{\mu\nu} + B'^J_{\mu\nu} Y_+(K'_J)^{\mu\nu} \right)
\end{equation}
where $K'_I$ and $K'_J$ are both $\A$- or $\C$-type CKY tensors and $B'^I$ and $B'^J$ are 2-form background gauge fields transforming as
\begin{equation}\label{eq:B''_background_transf}
    \delta B'^I_{\mu\nu} = \partial_{[\mu} \lambda'^I_{\nu]}\qc \delta B'^J_{\mu\nu} = \partial_{[\mu} \lambda'^J_{\nu]}
\end{equation}
The action \eqref{eq:4d_two_magnetic_gauged} is simultaneously invariant under both background transformations in \eqref{eq:B''_background_transf}. Therefore, there is no mixed 't Hooft anomaly between any pair of magnetic 1-form symmetries in four dimensions. One can then gauge any subgroup of the magnetic symmetries by introducing similar couplings.

Consider now $d>4$, where the magnetic symmetries correspond to the $(d-2)$-form conserved currents $\star Y(K'')$, where $K''$ is an $\A$- or $\B$-type CKY tensor. Each of these can be gauged by introducing the coupling shown in \eqref{eq:AC_coupling_d>4}. Just as in four dimensions, introducing multiple such couplings immediately gives an action which is invariant under the background gauge transformations of all the background fields. Explicitly, we consider the action
\begin{equation}
\begin{split}\label{eq:d>4_two_magnetic_gauged}
    S_2^{d>4} &= \frac{1}{2} \int \dd[d]{x} h_{\mu\nu} G^{\mu\nu} - \int \left( \tilde{B}^I \wedge Y(K''_I) + \tilde{B}^J \wedge Y(K''_J) \right) 
\end{split}
\end{equation}
where $K''_I$ and $K''_J$ are both $\A$- or $\B$-type CKY tensors and the 2-form background fields $\tilde{B}^I$ and $\tilde{B}^J$ transforms as
\begin{equation}\label{eq:Btilde'_background_transf}
    \delta \tilde{B}^I_{\mu_1\dots\mu_{d-2}} = \partial_{[\mu_1} \tilde{\lambda}^I_{\mu_2\dots\mu_{d-2}]}\qc \delta \tilde{B}^J_{\mu_1\dots\mu_{d-2}} = \partial_{[\mu_1} \tilde{\lambda}^J_{\mu_2\dots\mu_{d-2}]}
\end{equation}
Then the action \eqref{eq:d>4_two_magnetic_gauged} is simultaneously invariant under both background transformations in \eqref{eq:Btilde'_background_transf}, so there are no mixed 't Hooft anomalies between any pairs of magnetic $(d-3)$-form symmetries in $d>4$ dimensions. One can then gauge any subgroup of the magnetic symmetries by introducing similar couplings.

\section{Mixed 't Hooft anomalies with dual symmetries}
\label{sec:anomalies_with_duals}

Having discussed the gauging of the electric 1-form symmetries and the magnetic $(d-3)$-form symmetries in the previous sections, we now ask whether they can be gauged simultaneously, or whether there are mixed 't Hooft anomalies obstructing this.

\subsection{Four dimensions}
\label{sec:anomalies_with_duals_4d}

Recall that the electric 1-form symmetries correspond to $\B$- and $\D$-type CKY tensors, whereas the magnetic 1-form symmetries in four dimensions correspond to $\A$- and $\C$-type CKY tensors.

We consider the action \eqref{eq:gauged_action} where an electric 1-form symmetry labelled by a $\B$- or $\D$-type CKY tensor $K$ has been coupled to a background gauge field $B$. The action is gauge-invariant under \eqref{eq:electric_transf}. 

The ungauged graviton theory also has magnetic 1-form symmetries. Let us consider an arbitrary one labelled by an $\A$- or $\C$-type CKY tensor $K'$.
We now introduce another 2-form background gauge field $B'$, transforming as in \eqref{eq:B'_background_transf}, via the coupling shown in \eqref{eq:4d_AC_coupling} to gauge this symmetry.
The putative action with both background fields is
\begin{equation}
    S^{d=4}_4 = \int\dd[4]{x} \left( \frac{1}{2} h_{\mu\nu} G^{\mu\nu} - B_{\mu\nu} Y_+(K)^{\mu\nu} + f(K,B) + B'_{\mu\nu} Y_+(K')^{\mu\nu} \right)
\end{equation}
This is, indeed, gauge-invariant under \eqref{eq:B'_background_transf} as $Y_+(K')$ is identically conserved. However, the introduction of the final term implies that $S_4^{d=4}$ is no longer invariant under \eqref{eq:electric_transf}. 
This can be remedied by adding counter-terms involving $B$ and $B'$. One way to do so is to replace the $B'_{\mu\nu} Y_+(K')^{\mu\nu}$ with $B'_{\mu\nu} \mathcal{Y}_+(K')^{\mu\nu}$ where 
\begin{equation}
    \mathcal{Y}_+(K')_{\mu\nu} = 6 \delta_{\mu\nu\alpha\beta}^{\rho\sigma\gamma\delta} K'_{\rho\sigma} \mathcal{R'}\indices{^{\alpha\beta}_{\gamma\delta}}
\end{equation}
is defined in the same way as the improved Penrose 2-form \eqref{eq:improved_Penrose} with $R$ replaced by the improved Riemann tensor $\mathcal{R'}$ in \eqref{eq:modified_Riemann}. Since $\mathcal{R'}$ is invariant under \eqref{eq:electric_transf}, this term will not spoil the invariance of the action under \eqref{eq:electric_transf}.
The modified action is
\begin{equation}\label{eq:S5_4d_def}
    S_5^{d=4} = \int \dd[4]{x} \left( \frac{1}{2} h_{\mu\nu} G^{\mu\nu} - B_{\mu\nu} Y_+(K)^{\mu\nu} + f(K,B) + B'_{\mu\nu} \mathcal{Y}_+(K')^{\mu\nu} \right)
\end{equation}
Making this replacement is equivalent to inserting certain counter-terms involving $B$ and $B'$. 
However, we will now show that these counter-terms are not invariant under the $B'$ background gauge transformations \eqref{eq:B'_background_transf}, and so spoil this invariance of the action.
Under a $B'$ background transformation \eqref{eq:B'_background_transf}, $S_5^{d=4}$ varies by
\begin{equation}\label{eq:S5_4d_var}
    \delta S_5^{d=4} = \int\dd[4]{x} \partial_{[\mu} \lambda'_{\nu]} \mathcal{Y}_+(K')^{\mu\nu} 
\end{equation}
Using the definition of $\mathcal{R'}$ in \eqref{eq:modified_Riemann}, the variation can be rearranged into the form
\begin{equation}
\begin{split}
    \delta S_5^{d=4} &= 6 \int\dd[4]{x} \delta B'_{\mu\nu} \delta^{\mu\nu\alpha\beta}_{\rho\sigma\gamma\delta} \bigg( -(K'_{\alpha\beta}k^\gamma + 2 {K'}\indices{^\gamma_\alpha} k_\beta) \partial^{[\delta} B^{\rho\sigma]} - 2B\indices{_\alpha^\gamma} {K'}^{\rho\sigma} \partial^\delta k_\beta \\
    &\quad + B^{\gamma\delta} \tilde{K'}\indices{_\alpha^{\rho\sigma}} k_\beta - B_{\alpha\beta} \tilde{K'}^{\rho\sigma\gamma} k^\delta + B^{\gamma\delta} \tilde{K'}\indices{^\rho_{\alpha\beta}} k^\sigma + B^{\gamma\delta} K'_{\alpha\beta} \partial^\rho k^\sigma - 2B^{\gamma\delta} {K'}\indices{^\rho_\alpha} \partial^\sigma k_\beta \bigg)
\end{split}
\end{equation}
The first term inside the large parentheses depends only on $\dd{B}$ and so can be cancelled by introducing a counter-term
\begin{equation}\label{eq:Sct_4d_def}
    S_{\text{c.t.}}^{d=4} = 6 \int\dd[4]{x} B'_{\mu\nu} \delta^{\mu\nu\alpha\beta}_{\rho\sigma\gamma\delta}(K'_{\alpha\beta}k^\gamma + 2 {K'}\indices{^\gamma_\alpha} k_\beta) \partial^{[\delta} B^{\rho\sigma]}
\end{equation}
into the action. Since this depends only on $\dd{B}$, it does not transform under \eqref{eq:electric_transf}, so the new action 
\begin{equation}\label{eq:S6_4d}
    S_6^{d=4} = S_5^{d=4} + S_{\text{c.t.}}^{d=4}
\end{equation}
remains invariant under this transformation. The variation of $S_6^{d=4}$ under \eqref{eq:B'_background_transf} is then
\begin{equation}
\begin{split}\label{eq:S6_var_4d}
    \delta S_6^{d=4} &= 6 \int\dd[4]{x} \delta B'_{\mu\nu} \delta^{\mu\nu\alpha\beta}_{\rho\sigma\gamma\delta} \bigg( B^{\gamma\delta} \tilde{K'}\indices{_\alpha^{\rho\sigma}} k_\beta - B_{\alpha\beta} \tilde{K'}^{\rho\sigma\gamma} k^\delta + B^{\gamma\delta} \tilde{K'}\indices{^\rho_{\alpha\beta}} k^\sigma \\
    &\quad - 2B\indices{_\alpha^\gamma} {K'}^{\rho\sigma} \partial^\delta k_\beta + B^{\gamma\delta} K'_{\alpha\beta} \partial^\rho k^\sigma - 2B^{\gamma\delta} {K'}\indices{^\rho_\alpha} \partial^\sigma k_\beta \bigg)
\end{split}
\end{equation}
This can be simplified, using the fact that
$\delta^{\mu\nu\alpha\beta}_{\rho\sigma\gamma\delta}$   transforms in the [4,4] representation of $GL(d,\mathbb{R})$). The result can be written
\begin{equation}\label{eq:S6_var_4d_neater}
    \delta S_6^{d=4} = 6 \int \dd[4]{x} \delta B'_{\mu\nu} \delta^{\mu\nu\alpha\beta}_{\rho\sigma\gamma\delta} B_{\alpha\beta} \left( K'^{\rho\sigma} \partial^\gamma k^\delta - \frac{2}{3} \tilde{K}'^{\rho\sigma\gamma} k^\delta \right)
\end{equation}
This variation cannot be cancelled for general $k$ and $K'$ without spoiling invariance of the action under \eqref{eq:electric_transf}. However, for certain choices of $k$ and $K'$, it is possible to simultaneously couple background fields for both symmetries. This implies that there are mixed 't Hooft anomalies between certain electric 1-form symmetries and magnetic $(d-3)$-form symmetries. We show in Appendix~\ref{app:mixed_anomalies_4d} that there is a mixed 't Hooft anomaly when $k$ is a translational (i.e. constant) Killing vector and $K'$ is a $\C$-type CKY tensor, and also when $k$ is a rotational Killing vector and $K'$ is an $\A$-type CKY tensor. Otherwise, there is no mixed 't Hooft anomaly and the two background fields $B$ and $B'$ can both be consistently coupled. The different cases are summarised in Table~\ref{tab:4d_anomalies}. 

In the cases where there is an anomaly, the variation can be written in the form
\begin{equation}\label{eq:variation_general}
    \delta S_6^{d=4} = c \int \delta B' \wedge B
\end{equation}
where $c$ is a constant. This cannot be cancelled by a counter-term without spoiling invariance under \eqref{eq:electric_transf}. 
A nice way to see this is by the descent equations. The 5-dimensional anomaly theory which relates to this by descent is
\begin{equation}\label{eq:anomaly_theory}
    c \int B' \wedge \dd{B}
\end{equation}
which is a BF-type theory with $\mathbb{R}$-valued gauge fields. This is a non-trivial topological theory in 5-dimensions and so the variation \eqref{eq:variation_general} cannot be cancelled by a local 4-dimensional counter-term without spoiling invariance under the \eqref{eq:electric_transf} symmetry. The standard BF theory with $U(1)$ gauge fields has been discussed in \cite{Kapustin:2014gua, Maldacena:2001ss, Banks:2010zn}. Here the gauge fields are instead valued in $\mathbb{R}$. BF theories with non-compact gauge groups have been important recently in discussions of continuous non-invertible symmetry defects \cite{Arbalestrier:2024oqg} and the construction Symmetry Topological Field Theories (SymTFTs) for continuous global symmetries \cite{Brennan:2024fgj, Antinucci:2024zjp}.

\begin{table}
    \centering
    \begin{tabular}{c|c|c}
         $k$ & $K'$ & mixed 't Hooft anomaly? \\
         \hline
         translational & $\A$-type & no \\
         rotational & $\A$-type & yes \\
         translational & $\C$-type & yes \\
         rotational & $\C$-type & no
    \end{tabular}
    \caption{Mixed 't Hooft anomalies between electric and magnetic 1-form symmetries of the four-dimensional graviton. The `translational' Killing vectors are constant vectors, whereas the `rotational' ones are those linear in the coordinates, see \eqref{eq:Khat_solution}. See Appendix~\ref{app:mixed_anomalies_4d} for details.}
    \label{tab:4d_anomalies}
\end{table}

We note that, even in the cases where there are mixed 't Hooft anomalies between a pair of symmetries, these vanish when the background for the electric symmetry \eqref{eq:electric_transf} is flat, i.e. $\dd{B}=0$. This is immediately clear from the variation \eqref{eq:variation_general} which vanishes upon integration by parts in this case. Indeed, the anomaly theory \eqref{eq:anomaly_theory} vanishes in this case also.

The vanishing of the mixed 't Hooft anomalies when $\dd{B}=0$ can also be seen in the following way. Consider the action $S_5^{d=4}$ defined in \eqref{eq:S5_4d_def}. Its variation under the $B'$ background transformation \eqref{eq:B'_background_transf} is given in \eqref{eq:S5_4d_var}. Integrating by parts, this variation can be written
\begin{equation}\label{eq:S5_4d_var2}
    \delta S_5^{d=4} = - \int \dd[4]{x} \lambda'_\nu \partial_\mu \mathcal{Y}_+(K')^{\mu\nu}
\end{equation}
This can be evaluated using the properties of the improved Riemann tensor $\mathcal{R'}$ derived in section \ref{sec:gauging_Max}.
Explicitly, we find
\begin{align}
    \partial^\mu \mathcal{Y}_+(K')_{\mu\nu} &= 6 \delta_{\mu\nu\alpha\beta}^{\rho\sigma\gamma\delta} \partial^\mu \left( K'_{\rho\sigma} \mathcal{R'}\indices{^{\alpha\beta}_{\gamma\delta}} \right) \nonumber \\
    &= 6 \delta_{\mu\nu\alpha\beta}^{\rho\sigma\gamma\delta} \left( \tilde{K'}\indices{^\mu_{\rho\sigma}} + 2 \delta^\mu_{[\rho} \hat{K'}_{\sigma]} \right) \mathcal{R'}\indices{^{\alpha\beta}_{\gamma\delta}} + 6 \delta_{\mu\nu\alpha\beta}^{\rho\sigma\gamma\delta} K'_{\rho\sigma} \partial^{[\mu} \mathcal{R'}\indices{^{\alpha\beta]}_{\gamma\delta}} \nonumber \\
    &= 4 \delta_{\nu\alpha\beta\gamma}^{\delta\mu\rho\sigma} \tilde{K'}_{\mu\rho\sigma} \mathcal{R'}\indices{^{[\alpha\beta\gamma]}_\delta} + 6 \delta_{\mu\nu\alpha\beta}^{\rho\sigma\gamma\delta} K'_{\rho\sigma} \partial^{[\mu} \mathcal{R'}\indices{^{\alpha\beta]}_{\gamma\delta}} \label{eq:div_curly_Y}
\end{align}
where we have used the CKY equation \eqref{eq:CKY_equation} in the second equality and the fact that $\eta_{\nu\alpha\beta[\mu|\rho\sigma\gamma\delta]}=0$ in the third. We have also used that $\hat{K}'=0$, since $K'$ is an $\A$- or $\C$-type CKY tensor.
From \eqref{eq:gauged_non_bianchi3} and \eqref{eq:gauged_non_bianchi4}, the right-hand side of \eqref{eq:div_curly_Y} does not vanish in general.
However, if we restrict $B$ to be a flat background field configuration, so $\dd{B}=0$, then \eqref{eq:gauged_non_bianchi3} and \eqref{eq:gauged_non_bianchi4} imply that the relevant Bianchi identities and field equations are satisfied such that $\mathcal{Y}_+(K')$ is conserved and the variation \eqref{eq:S5_4d_var2} vanishes. Now, from \eqref{eq:S6_4d} we note that $S_5^{d=4}$ and $S_6^{d=4}$ differ by $S_{\text{c.t.}}^{d=4}$, which depends on $B$ only through its curvature $\dd{B}$ (see \eqref{eq:Sct_4d_def}). So the variation of $S_6^{d=4}$ under the $B'$ background transformations will also vanish when $\dd{B}=0$.

\subsection{\texorpdfstring{$d>4$}{d>4} dimensions}
\label{sec:anomalies_with_duals_d>4}

In $d>4$ dimensions, there are electric 1-form symmetries generated by Penrose charges corresponding to $\B$- and $\D$-type CKY tensors, $K$. These can be gauged as in \eqref{eq:gauged_action} with a 2-form background gauge field $B$. There are also an equal number of magnetic $(d-3)$-form symmetries, labelled by $\A$- and $\B$-type CKY tensors, $K''$. These can be gauged as in \eqref{eq:AC_coupling_d>4} with a $(d-2)$-form background gauge field $\tilde{B}$. 
We again begin with the action \eqref{eq:gauged_action}, which has a gauge symmetry \eqref{eq:electric_transf}, and introduce a $(d-2)$-form background field $\tilde{B}$ via the coupling
\begin{equation}
    S_4^{d>4} = \int\dd[d]{x} \left( \frac{1}{2} h_{\mu\nu} G^{\mu\nu} - B_{\mu\nu} Y_+(K)^{\mu\nu} + f(K,B) + \frac{1}{(d-2)!} \tilde{B}_{\mu_1\dots\mu_{d-2}} (\star Y(K''))^{\mu_1\dots\mu_{d-2}} \right) 
\end{equation}
This is invariant under $\tilde{B}$ background transformations \eqref{eq:Btilde_background_transf}, but not under \eqref{eq:electric_transf} since the $\tilde{B} \star Y(K'')$ term transforms non-trivially. In order to restore this invariance, we can simply replace $R_{\mu\nu\rho\sigma}$ by $\mathcal{R'}_{\mu\nu\rho\sigma}$ in the definition of $Y(K'')^{\mu\nu}$ in the final term, since $\mathcal{R'}$ in \eqref{eq:modified_Riemann} is invariant under \eqref{eq:electric_transf}. This modified action can then be written
\begin{equation}\label{eq:S5_d>4_def}
    S_5^{d>4} = \int\dd[d]{x} \left( \frac{1}{2} h_{\mu\nu} G^{\mu\nu} - B_{\mu\nu} Y_+(K)^{\mu\nu} + f(K,B) + \frac{1}{2} (\star \tilde{B})^{\mu\nu} \mathcal{R'}_{\mu\nu\alpha\beta} {K''}^{\alpha\beta} \right)
\end{equation}
and is invariant under \eqref{eq:electric_transf}.
However, it is no longer invariant under the $\tilde{B}$ background transformation \eqref{eq:Btilde_background_transf}. 
The variation of the action \eqref{eq:S5_d>4_def} under \eqref{eq:Btilde_background_transf} is
\begin{align}
    \delta S_5^{d>4} &= \frac{1}{2} \int \dd[d]{x} (\star \delta\tilde{B})^{\mu\nu} \mathcal{R'}_{\mu\nu\alpha\beta} K''^{\alpha\beta} \label{eq:S5_d>4_var0} 
\end{align}
Substituting in $\mathcal{R}'$ as defined in \eqref{eq:modified_Riemann} and integrating by parts, this can be written
\begin{equation}
    \delta S_5^{d>4}=  -\frac{3}{2} \int \dd[d]{x} (\star \delta\tilde{B})^{\mu\nu} K''_{\alpha\beta} k^\alpha (\dd{B})\indices{^\beta_{\mu\nu}} + \frac{1}{2} \int \dd[d]{x} (\star \delta\tilde{B})^{\mu\nu} B_{\mu\nu} \left( K''_{\alpha\beta} \partial^\alpha k^\beta - 2 \hat{K}''_{\alpha} k^\alpha \right)
\end{equation}
We note that the first term depends only on $\dd{B}$, and so can be cancelled by introducing a counter-term 
\begin{equation}
    S_{\text{c.t.}}^{d>4} = \frac{3}{2} \int \dd[d]{x} (\star \tilde{B})^{\mu\nu} K''_{\alpha\beta} k^\alpha (\dd{B})\indices{^\beta_{\mu\nu}}
\end{equation}
into the action. Since this depends only on $\dd{B}$, it is invariant under \eqref{eq:electric_transf}, and so the new action
\begin{equation}\label{eq:S6_d>4_def}
    S_6^{d>4} = S_5^{d>4} + S_{\text{c.t.}}^{d>4}
\end{equation}
is still invariant under this symmetry. The variation of $S_6^{d>4}$ under the $\tilde{B}$ background transformation \eqref{eq:Btilde_background_transf} is then
\begin{equation}\label{eq:S6_d>4_var}
    \delta S_6^{d>4} = \frac{1}{2} \int \dd[d]{x} (\star \delta\tilde{B})^{\mu\nu} B_{\mu\nu} \left( K''_{\alpha\beta} \partial^\alpha k^\beta - 2 \hat{K}''_{\alpha} k^\alpha \right)
\end{equation}

We show in Appendix~\ref{app:mixed_anomalies_d>4} that only for certain choices of $K$ and $K''$ can counter-terms be found which make the action invariant under both \eqref{eq:electric_transf} and \eqref{eq:Btilde_background_transf} simultaneously. When this is not possible, there is a mixed 't Hooft anomaly between the electric 1-form and magnetic $(d-3)$-form symmetries. We find that there is a mixed 't Hooft anomaly when $k$ is a translational (i.e. constant) Killing vector and $K''$ is a $\B$-type CKY tensor, and also when $k$ is a rotational Killing vectors and $K''$ is an $\A$-type CKY tensor. Otherwise, there is no mixed anomaly. A summary of the results is given in Table~\ref{tab:d>4_anomalies}.

When there is a mixed 't Hooft anomaly, the variation of $S_6^{d>4}$ can be written in the form \eqref{eq:variation_general} and so, as in the discussion in the previous subsection, cannot be cancelled by any local $d$-dimensional counter-term without spoiling invariance under \eqref{eq:electric_transf}.

\begin{table}
    \centering
    \begin{tabular}{c|c|c}
         $k$ & $K''$ & mixed 't Hooft anomaly? \\
         \hline
         translational & $\A$-type & no \\
         rotational & $\A$-type & yes \\
         translational & $\B$-type & yes \\
         rotational & $\B$-type & no
    \end{tabular}
    \caption{Mixed 't Hooft anomalies between electric and magnetic 1-form symmetries of the $d$-dimensional graviton for $d>4$. The `translational' Killing vectors are constant vectors, whereas the `rotational' ones are those linear in the coordinates, see \eqref{eq:Khat_solution}. See Appendix~\ref{app:mixed_anomalies_d>4} for details.}
    \label{tab:d>4_anomalies}
\end{table}

Even in the cases where there are mixed 't Hooft anomalies, these are proportional to $\dd{B}$ and so vanish when the background of the electric 1-form symmetry \eqref{eq:electric_transf} is flat, i.e. $\dd{B}=0$. This can also be discerned by examining the variation \eqref{eq:S5_d>4_var0}, which can be written
\begin{align}
    \delta S_5^{d>4} &= \frac{1}{2(d-2)!} \int \dd[d]{x} \epsilon^{\mu\nu\gamma_1\dots\gamma_{d-2}} \partial_{\gamma_1} \tilde{\lambda}_{\gamma_2 \dots\gamma_{d-2}}  \mathcal{R'}_{\mu\nu\alpha\beta} K''^{\alpha\beta} \nonumber \\
    &= - \frac{1}{2(d-2)!} \int \dd[d]{x} \epsilon^{\mu\nu\gamma_1\dots\gamma_{d-2}} \tilde{\lambda}_{\gamma_2 \dots\gamma_{d-2}} \left( \partial_{[\gamma_1} \mathcal{R'}_{\mu\nu]\alpha\beta} K''^{\alpha\beta} + 2 \mathcal{R'}_{[\mu\nu\gamma_1]\beta} \hat{K}''^\beta \right)
\end{align}
where we have used the form of the variation \eqref{eq:Btilde_background_transf} in the first equality and the fact that $K''$ is an $\A$- or $\B$-type CKY tensor in the second. While \eqref{eq:gauged_non_bianchi3} and \eqref{eq:gauged_non_bianchi4} show that this variation does not vanish in general, they do imply that it vanishes when $\dd{B}=0$, in agreement with our results above.

The cases where 't Hooft anomalies occur between the Penrose charges $Q(K)$ and the dual Penrose charges match the discussion of dualities between the Penrose charges in \cite{Hull:2024xgo}. 
Recall that in four dimensions, the dual charges are the Penrose charges $Q(K')$ where $K'$ is an $\A$- or $\C$-type CKY tensor. In $d>4$, the dual charges are the $\tilde{Q}(K'')$ defined in \eqref{eq:dual_Penrose_charges} where $K''$ is an $\A$- or $\B$-type CKY tensor.
In particular, it was found in that work that in four dimensions the $\A$- and $\D$-type Penrose charges can be viewed as dual, and similarly for the $\B$- and $\C$-type charges. In $d>4$ dimensions, the $\B$-type Penrose charges are dual to the $\B$-type dual Penrose charges, and the $\D$-type Penrose charges are dual to the $\A$-type dual charges. These combinations agree with those summarised in Tables~\ref{tab:4d_anomalies} and \ref{tab:d>4_anomalies}.
Furthermore, in \cite{Benedetti2023GeneralizedGravitons}, the commutator between the Penrose charges $Q(K)$ and the dual Penrose charges $\tilde{Q}(K'')$ was computed using canonical quantisation of the graviton theory in $d>4$. It was found that the only non-vanishing commutators are between the $\B$-type Penrose charges and the $\B$-type duals, and between the $\D$-type Penrose charges and the $\A$-type duals. Again, this matches the combinations of charges exhibiting mixed 't Hooft anomalies.

\subsection{Significance of the anomalies for linearised gravity}

We have seen that a pair of dual generalised symmetries of the linearised graviton theory have mixed anomalies. This has been analysed as an obstruction to simultaneously gauging the dual symmetries. We now discuss the implications of this for the original (ungauged) linearised graviton theory. The arguments are a straightforward extension of those in 
\cite{Hinterbichler2023GravitySymmetries,Delacretaz:2019brr} and can be summarized as follows.

In the ungauged theory, there is a  current corresponding to each of the two dual symmetries. Of particular interest is the 2-point correlation function of these two currents. The anomaly leads to the 2-point function satisfying an anomalous Ward identity, corresponding to one of the two currents no longer being conserved.  This in turn fixes the singular terms  in the correlator. For any theory which has currents with such a correlator,   the   singular terms should be invariant under renormalisation group flow. In particular, the massless pole indicates the existence of a Goldstone mode.

\section{Comparing shift symmetries of the graviton}
\label{sec:comparison_with_gapless}

In \cite{Hinterbichler2023GravitySymmetries}, shifts of the form
\begin{equation}\label{eq:dh=deltaLambda}
    \delta h_{\mu\nu} = \partial^\rho \Lambda_{\rho(\mu|\nu)}
\end{equation}
were considered, where $\Lambda$ is a tensor field in the [2,1] representation of $GL(d,\mathbb{R})$ satisfying
\begin{equation}
    G(\delta h)=0
\end{equation}
where $G$ is the Einstein tensor defined in \eqref{eq:G(alpha)=0}. The [2,2] bi-form Noether current which is associated with this symmetry can be taken as the linearised Riemann tensor $R$. Note that a term of the form $\partial^\sigma \Delta_{\sigma\rho\mu|\nu}$, where $\Delta$ is a [3,1] tensor, can be added to $\Lambda_{\rho\mu|\nu}$ without changing the transformation $\delta h$. The fact that the Penrose 2-form $Y(K)$ and its improvement $Y_+(K)$ are contractions of $R$ with a CKY tensor $K$ hints at a relation between the global symmetries \eqref{eq:dh=deltaLambda} and \eqref{eq:dh=lambda_k}. 

Consider $\Lambda$ of the form
\begin{equation}\label{eq:Lambda=Klambda}
    \Lambda_{\rho\mu|\nu} = K_{\rho\mu}\lambda_\nu - K_{[\rho\mu}\lambda_{\nu]} - \frac{2}{d-1} \eta_{\nu [\mu} K_{\rho]\sigma} \lambda^\sigma
\end{equation}
where, as in previous sections, $K$ is a $\B$- or $\D$-type CKY tensor and $\lambda$ is a closed 1-form. In this case, the resulting transformation of the graviton field is 
\begin{equation}\label{eq:dh_Lambda}
    \delta h_{\mu\nu} = \partial^\rho \Lambda_{\rho(\mu|\nu)} = \frac{d-2}{2(d-3)} \lambda_{(\mu} k_{\nu)} - \frac{d-2}{d-1} \partial_{(\mu} \left( K_{\nu)\rho} \lambda^\rho \right)
\end{equation}
where $k$ is related to $K$ by \eqref{eq:k=Khat} as before. 
The first term on the right-hand side of \eqref{eq:dh_Lambda} is of the form \eqref{eq:dh=lambda_k} which, from section~\ref{sec:particular_shifts}, is a global symmetry since $k$ is a Killing vector and $\dd{\lambda}=0$. As discussed previously, this transformation relates to the Penrose charges. The final term in \eqref{eq:dh_Lambda} is a linearised diffeomorphism and so is a gauge symmetry of the theory for any $K$ and $\lambda$.

Therefore, we see that the transformations \eqref{eq:dh=lambda_k} are a subcase of the transformations \eqref{eq:dh_Lambda} where $\Lambda$ is given by \eqref{eq:Lambda=Klambda}. The most general shift symmetries of the graviton, $\delta h_{\mu\nu} = \alpha_{\mu\nu}$, were studied in \cite{Hull:2024bcl}. These are global symmetries if $\alpha$ is a [1,1] tensor satisfying the flatness condition $G(\alpha)=0$ (that is, the linearised Einstein tensor built from $\alpha$ vanishes). Clearly, the shifts \eqref{eq:dh_Lambda} are a subcase of these more general symmetries. 

This nested structure of the shift symmetries is reflected in the structure of the Noether currents associated with each, as we now describe. The Noether current associated with the most general shifts $\delta h = \alpha$ is a [2,1] tensor with components \cite{Hull:2024bcl}
\begin{equation}
    J\indices{_{\mu\nu|}^\rho} = 3 \delta_{\mu\nu\alpha}^{\rho\sigma\beta} \Gamma\indices{_{\sigma\beta|}^\alpha}
\end{equation}
which has the property that 
\begin{equation}
    \partial^\mu J_{\mu\nu|\rho} = G_{\mu\nu}
\end{equation}
so is conserved on-shell. This Noether current is not gauge-invariant, which leads to several subtleties in the discussion of the corresponding charges \cite{Hull:2024bcl}. Now consider the symmetries \eqref{eq:dh_Lambda}, which are a subcase of the former, whose associated Noether current is the [2,2] tensor $R_{\mu\nu\rho\sigma}$. It was shown in \cite{Hull:2024bcl} that the on-shell conservation of $R$ can be seen as a consequence of the on-shell conservation of $J$. Next, we have seen that the symmetries \eqref{eq:dh=lambda_k} are a subcase of \eqref{eq:dh_Lambda}, with Noether current $Y_+(K)$. On-shell, $Y_+(K) = Y(K)$ and so its divergence is
\begin{equation}
    \partial^\mu Y_+(K)_{\mu\nu} = \partial^\mu Y(K)_{\mu\nu} = \partial^\mu R_{\mu\nu\rho\sigma} K^{\rho\sigma} + R_{\mu\nu\rho\sigma} \left( \tilde{K}^{\mu\rho\sigma} + 2\eta^{\mu\rho} \hat{K}^\sigma \right)
\end{equation}
This vanishes from the on-shell conservation of $R$ and its [2,2] tensor structure. Therefore, we see that the on-shell conservation of the Noether currents associated with each sub-symmetry follows from the conservation of the current associated with the more general shift symmetries, as one would expect. 
\section{Discussion \& outlook}
\label{sec:conclusion}

The higher-form global symmetries of the graviton theory   differ in several interesting ways from those of $p$-form gauge theory which, starting with \cite{Gaiotto2015GeneralizedSymmetries}, have been very well explored in the literature. 

Firstly, the structure of the 1-form and $(d-3)$-form symmetries of the graviton is closely tied to the properties of conformal Killing-Yano tensors. 
The electric 1-form symmetries of the graviton can be interpreted as stemming from Noether currents for  particular global shift symmetries of the graviton field, while the magnetic $(d-3)$-form symmetries are not related to any  transformation of the graviton. The electric symmetries were gauged via the Noether  coupling to a 2-form background gauge field, $B$, and finding counter-terms   quadratic in $B$. 
The magnetic symmetries were gauged instead by coupling the conserved current to a $(d-2)$-form background gauge field.

The electric symmetries do not have 't Hooft anomalies, nor are there any mixed 't Hooft anomalies between them. The same is true of the magnetic symmetries. However, there is an interesting set of mixed 't Hooft anomalies between certain pairs of electric and magnetic symmetries. 
We found the full set of anomalies and showed how they can be described by descent equations, discussing the cases $d=4$ and $d>4$ separately.
 Our results agree with indirect evidence found in \cite{Benedetti2023GeneralizedGravitons} where the commutation relations between the charges generating the higher-form symmetries were studied.

It is interesting to contrast the higher-form symmetry structure found in linear gravity with that of Maxwell theory. In Maxwell theory, the 1-form symmetry stems from the conservation of $F=\dd{A}$. The field equation and Bianchi identity are, respectively,
\begin{equation}\label{eq:Maxwell_eqs}
    \dd^\dag F=0\qc \dd{F} = 0
\end{equation}
In contrast, the 1-form symmetries of the graviton are generated by the Penrose 2-form $Y(K)$ where $K$ is a CKY 2-form. In particular, this requires the existence of CKY tensors on the background spacetime, which here is Minkowski space. As pointed out in \cite{Hinterbichler2023GravitySymmetries, Hull:2024bcl}, the linearised Riemann tensor itself $R_{\mu\nu\rho\sigma}$ can be thought of as a conserved  current since it follows from the Bianchi identity $\partial_{[\mu} R_{\nu\rho]\alpha\beta}=0$ that $\partial^\mu R_{\mu\nu\rho\sigma}=0$ on-shell. In the language of \cite{Medeiros2003ExoticDuality}, $R$ is a [2,2] bi-form.
Then each  CKY tensor $K$ can then be thought of providing a  projection from the [2,2] current $R$ to the 2-form current $Y_+(K)$. In other words, the different CKY tensors $K$ give different ways to isolate a spin-1 substructure within the spin-2 structure encoded in the Riemann tensor $R$.

Furthermore, for some CKY tensors, $Y(K)$ is both co-closed and closed on-shell. In $d=4$ this is the case for all CKY tensors, while in $d>4$ only the closed CKY tensors have this property. Of those, only the $\B$-type tensors give rise to non-vanishing Penrose charges $Q(K)$ and dual Penrose charges $\tilde{Q}(K)$ (see sections~\ref{sec:Penrose_review} and \ref{sec:dual_symms}). If $K$ is any one of these CKY tensors, then $F=Y(K)$ solves both of the Maxwell equations \eqref{eq:Maxwell_eqs} on-shell. This gives a map from solutions of the graviton theory to solutions of Maxwell theory. More specifically, consider a solution of the graviton field equation and Bianchi identities
\begin{equation}
    R\indices{^\mu_{\nu\mu\rho}} = 0 \qc R_{[\mu\nu\rho]\sigma} = 0\qc \partial_{[\mu} R_{\nu\rho]\alpha\beta} = 0
\end{equation}
Then $K$ gives a map $R_{\mu\nu\rho\sigma} \to Y(K)_{\mu\nu} = R_{\mu\nu\alpha\beta} K^{\alpha\beta}$ from solutions of the linearised gravity theory to solutions of Maxwell theory. One can verify, for example, that the linearised Schwarzschild solution maps to a Coulomb potential in this manner, which is somewhat reminiscent of the Kerr-Schild \cite{Monteiro:2014cda} and, more generally, Weyl double copies \cite{Luna:2018dpt}. It would be interesting to study whether this behaviour can generalise to other background spacetimes.

The graviton theory also has a dual formulation \cite{Hull2000} in terms of a tensor gauge field in the $[d-3,1]$ representation of $GL(d,\mathbb{R})$. The dual description must have the same global symmetries as the original formulation of the theory. The mapping of the electric and magnetic symmetries to their counterparts in the dual theory is an interesting correspondence which we will return to in upcoming work. 
The duality between the higher-form symmetries in the graviton theory is far richer than in pure $p$-form gauge theory where there is a single pair of electric and magnetic charges which are interchanged under duality.
Furthermore, it would be interesting to study the generalisation of the results found here to gauge fields in arbitrary representations of $GL(d,\mathbb{R})$, in particular fully symmetric higher-spin gauge fields.

Maxwell theory in four dimensions has `duality defects' \cite{Choi:2021kmx, Choi:2022zal}, which are non-invertible topological operators (see also \cite{Kaidi:2021xfk}, and \cite{Shao:2023gho} for a  review).
At particular values of the coupling, gauging a $\mathbb{Z}_N$ subgroup of  the electric 1-form symmetry and applying an S-duality transformation maps Maxwell theory back to itself. The defect is constructed by performing this operation in half of  spacetime, which then defines a defect at the boundary between the two halves. An analogous construction is possible for a compact boson in two dimensions for particular values of the radius \cite{Thorngren:2021yso,Choi:2021kmx}. In that case, the fusion rules of the duality defect and the generator of the $\mathbb{Z}_N$ subgroup of the 1-form symmetry are then given by the $\mathbb{Z}_N$ Tambara-Yamagami category \cite{Tambara:1998vmj}. (See \cite{Argurio:2024ewp} for a recent generalisation of this construction to non-invertible defects at any radius of the compact boson). Given the rich structure of  1-form symmetries of the 4-dimensional graviton theory found here, it is natural to ask whether such duality defects arise also in the spin-2 case. We hope to return to this question in future work.\footnote{We are grateful to the referee for suggesting this interesting possibility.}

Finally, while the linear theory of the graviton has a surprisingly rich symmetry structure, it is important to understand how much of it carries over to Einstein gravity. 
The relation between transformations of the form \eqref{eq:dh=lambda_k} and conformal rescalings has been explored for the non-linear Einstein-Hilbert action in \cite{Gomez-Fayren:2024cpl}, with some extra constraints on $\lambda$ and $k$ being necessary for the transformation to be a symmetry of the non-linear action.
Only certain rather special spacetimes admit conformal Killing-Yano tensors and  for these  there are analogues of the symmetries considered here.
General spacetimes admit no CKY tensors and so have no generalised symmetries of the kind considered here, but on spacetimes that are asymptotic to ones admitting such tensors there can be asymptotic generalised symmetries.
We will return these and related questions in future work.

\paragraph{Acknowledgements.}
CH is supported by the STFC Consolidated Grants ST/T000791/1 and ST/X000575/1.
MLH is supported by a President's Scholarship from Imperial College London.
UL gratefully acknowledges a Leverhulme Visiting Professorship to Imperial College as well as the hospitality of the theory group at Imperial. 

\appendix
\section{Determining counter-terms for gauging 1-form symmetries}
\label{app:counterterms}

In this appendix we give a systematic derivation of the counter-terms $f(B)$ required to gauge the shift symmetry \eqref{eq:dh=lambda_k}, such that the action in \eqref{eq:gauged_action} is gauge-invariant. We recall that the global shift symmetry to be gauged is
\begin{equation}
    \delta h_{\mu\nu} = \lambda_{(\mu} k_{\nu)}
\end{equation}
where $\dd\lambda=0$ and $k$ is a Minkowski space Killing vector. Gauging this symmetry will result in a theory which has this as a symmetry without the constraint on $\lambda$. This is a 1-form symmetry and so a 2-form gauge field $B$ transforming as
\begin{equation}
    \delta B_{\mu\nu} = \partial_{[\mu} \lambda_{\nu]}
\end{equation}
is the relevant background field to couple. The improved Penrose current $Y_+(K)$ can be understood as the Noether current for this symmetry and so the gauging should be accomplished by introducing couplings to $B$ as in \eqref{eq:gauged_action}, which we repeat here for convenience:
\begin{equation}
    S'' = \int\dd[d]{x} \left( \frac{1}{2} h_{\mu\nu} G^{\mu\nu} - B_{\mu\nu} Y_+(K)^{\mu\nu} + f(K,B) \right)
\end{equation}
The variation of the graviton action is cancelled by the variation of $B$ in the second term. The variation of $h$ in the second term (recall that $Y_+(K)$ is a function of $h$) must then cancel against the variation of $f(K,B)$, which is a quadratic function of $B$. 

\subsection{Variation of the Penrose current}

We begin by identifying the variation of the second term. The improved Penrose 2-form can be written in the convenient form
\begin{equation}\label{eq:Y+_nice_form}
    Y_+(K)^{\mu\nu} = 6 \delta^{\mu\nu\alpha\beta}_{\rho\sigma\gamma\delta} K^{\rho\sigma} R\indices{_{\alpha\beta}^{\gamma\delta}} = -12 \delta^{\mu\nu\alpha\beta}_{\rho\sigma\gamma\delta} K^{\rho\sigma} \partial_\alpha \partial^\gamma h\indices{_\beta^\delta}
\end{equation}
such that the variation in question is
\begin{align}
    -B_{\mu\nu} \delta Y_+(K)^{\mu\nu} &= 12 \delta^{\mu\nu\alpha\beta}_{\rho\sigma\gamma\delta} B_{\mu\nu} K^{\rho\sigma} \partial_\alpha \partial^\gamma \delta h\indices{_\beta^\delta} \nonumber \\
    &= 6 \delta^{\mu\nu\alpha\beta}_{\rho\sigma\gamma\delta} B_{\mu\nu} K^{\rho\sigma} \big( \partial^\gamma \partial_{[\alpha} \lambda_{\beta]} k^\delta + \partial_{[\alpha} \lambda_{\beta]} \partial^\gamma k^\delta + \partial^\gamma \lambda_\beta \partial_\alpha k^\delta \nonumber \\
    & \qquad \qquad \qquad \quad + \partial_\alpha \partial^{[\gamma} \lambda^{\delta]} k_\beta + \partial^{[\gamma} \lambda^{\delta]} \partial_\alpha k_\beta + \partial_\alpha \lambda^\delta \partial^\gamma k_\beta \big) \label{eq:BdY_working}
\end{align}
where we have used $\partial_\mu \partial_\nu k_\alpha=0$.
There are two terms in \eqref{eq:BdY_working} with two derivatives acting on $\lambda$. We can integrate these terms by parts to yield contributions of the form $\partial_{[\mu}\lambda_{\nu]} X^{\mu\nu}$. There are another two terms which are already of this form. The remaining two terms can be rearranged using the symmetries of the generalised Kronecker delta function and the fact that $k$ is a Killing vector, so $\partial_\mu k_\nu = \partial_{[\mu} k_{\nu]}$, to give
\begin{align}
\begin{split}
    -B_{\mu\nu} \delta Y_+(K)^{\mu\nu} = 6 \delta^{\mu\nu\alpha\beta}_{\rho\sigma\gamma\delta} \big( &- \partial_{[\alpha} \lambda_{\beta]} \partial^\gamma (B_{\mu\nu} K^{\rho\sigma} k^\delta) - \partial^{[\gamma} \lambda^{\delta]} \partial_\alpha (B_{\mu\nu} K^{\rho\sigma} k_\beta ) \\
    & + \partial_{[\alpha} \lambda_{\beta]} B_{\mu\nu} K^{\rho\sigma} \partial^\gamma k^\delta + \partial^{[\gamma} \lambda^{\delta]} B_{\mu\nu} K^{\rho\sigma} \partial_\alpha k_\beta \\
    & + (\partial^\gamma \lambda_\beta - \partial_\beta \lambda^\gamma) B_{\mu\nu} K^{\rho\sigma} \partial_\alpha k^\delta \big) 
\end{split}
\end{align}
Then expanding the generalised Kronecker delta function and relabeling indices yields
\begin{align}
    -B_{\mu\nu} \delta Y_+(K)^{\mu\nu} 
    &= -12 (d-3) \delta B_{\alpha\beta} \Big( \partial^{[\alpha} B_{\mu\nu} K^{\mu\nu} \hat{K}^{\beta]} + B_{\mu\nu} \tilde{K}^{[\alpha\mu\nu} \hat{K}^{\beta]} \nonumber \\
    &\quad + \partial^{[\alpha} B^{\mu\nu} K_{\mu\nu} \hat{K}^{\beta]} + B^{[\mu\nu} \partial^\alpha K_{\mu\nu} \hat{K}^{\beta]} + 2 \eta^{\beta[\alpha} B^{\rho\sigma} K_{\rho\sigma} \partial^{\delta]} \hat{K}_\delta \Big) \label{eq:BdY_variation_original}
\end{align}
where we have used \eqref{eq:k=Khat}. This is the variation that must be cancelled by the variation of $f(K,B)$.

\subsection{The set of all possible counter-terms}

Given the objects involved in each term of \eqref{eq:BdY_variation_original}, $f(K,B)$ must be constructed from terms involving two factors of $B$, a derivative $\partial$, a factor of $K$ and a factor of $\hat{K}$. By integrating by parts, we can always remove all derivatives from one factor of $B$ and write $f(K,B)$ as
\begin{equation}\label{eq:f=BF}
    f(K,B) = B_{\mu\nu} F(K,B)^{\mu\nu}
\end{equation}
where $F(K,B)$ is a 2-form linear in $B$ and also involves a derivative, a factor of $K$ and a factor of $\hat{K}$. It is non-trivial to write down an $F(K,B)$ such that $\delta f(B)$ cancels \eqref{eq:BdY_variation_original} so we will take a systematic approach. We begin by writing down all 2-forms which can be made from the relevant objects. We will then study their variations and show that $F(K,B)$ can be determined uniquely.

There are 21 independent 2-forms that can be constructed. We break them into three types. Firstly, those where the derivative acts on $B$,
\begin{align}
    g^1_{\mu\nu} &= \partial_\alpha B^{\alpha\beta} K_{\beta[\mu} \hat{K}_{\nu]} \label{eq:g1} \\
    g^2_{\mu\nu} &= \partial^\alpha B_{\alpha[\mu} K_{\nu]\beta} \hat{K}^\beta \label{eq:g2} \\
    g^3_{\mu\nu} &= \partial^\alpha B_{\alpha\beta} K_{\mu\nu} \hat{K}^\beta \label{eq:g3} \\
    g^4_{\mu\nu} &= \partial_\alpha B_{\mu\nu} K^{\alpha\beta} \hat{K}_\beta \label{eq:g4} \\
    g^5_{\mu\nu} &= \partial_\alpha B_{\beta[\mu} K^{\alpha\beta} \hat{K}_{\nu]} \label{eq:g5} \\
    g^6_{\mu\nu} &= \partial_\alpha B_{\beta[\mu} K\indices{_{\nu]}^\alpha} \hat{K}^\beta \label{eq:g6} \\
    g^7_{\mu\nu} &= \partial_\alpha B_{\beta[\mu} K\indices{_{\nu]}^\beta} \hat{K}^\alpha \label{eq:g7} \\
    g^8_{\mu\nu} &= \partial_{[\mu|} B_{\alpha\beta} K^{\alpha\beta} \hat{K}_{|\nu]} \label{eq:g8} \\
    g^9_{\mu\nu} &= \partial_{[\mu} B_{\nu]\alpha} K^{\alpha\beta} \hat{K}_\beta \label{eq:g9} \\
    g^{10}_{\mu\nu} &= \partial_{[\mu} B^{\alpha\beta} K_{\nu]\alpha} \hat{K}_\beta \label{eq:g10}
\end{align}
Secondly, those where the derivative acts on the CKY tensor. These can be expanded using the CKY equation to give some terms involving $\tilde{K}$ and some involving $\hat{K}$. A complete set is
\begin{align}
    p^1_{\mu\nu} &= B^{\alpha\beta} \tilde{K}_{\alpha\beta[\mu} \hat{K}_{\nu]} \label{eq:p1} \\
    p^2_{\mu\nu} &= B\indices{^\alpha_{[\mu}} \tilde{K}_{\nu]\alpha\beta} \hat{K}^\beta \label{eq:p2} \\
    p^3_{\mu\nu} &= B_{\alpha\beta} \tilde{K}\indices{_{\mu\nu}^\alpha} \hat{K}^\beta \label{eq:p3} \\
    p^4_{\mu\nu} &= B_{\mu\nu} \hat{K}_\alpha \hat{K}^\alpha \label{eq:p4} \\
    p^5_{\mu\nu} &= B_{\alpha[\mu} \hat{K}_{\nu]} \hat{K}^\alpha \label{eq:p5} 
\end{align}
Finally, those where the derivative acts on $\hat{K}$,
\begin{align}
    q^1_{\mu\nu} &= B_{\alpha\beta} K\indices{^\beta_{[\mu}} \partial_{\nu]} \hat{K}^\alpha \label{eq:q1} \\
    q^2_{\mu\nu} &= B_{\alpha[\mu} K_{\nu]\beta} \partial^\alpha \hat{K}^\beta \label{eq:q2} \\
    q^3_{\mu\nu} &= B_{\alpha\beta} K_{\mu\nu} \partial^\alpha \hat{K}^\beta \label{eq:q3} \\
    q^4_{\mu\nu} &= B_{\mu\nu} K_{\alpha\beta} \partial^\alpha \hat{K}^\beta \label{eq:q4} \\
    q^5_{\mu\nu} &= B_{\beta[\mu} K^{\alpha\beta} \partial_{\nu]} \hat{K}_\alpha \label{eq:q5} \\
    q^6_{\mu\nu} &= B_{\alpha\beta} K^{\alpha\beta} \partial_\mu \hat{K}_\nu \label{eq:q6} 
\end{align}

While all 21 two-forms listed above are independent of each other, when contracted with $B_{\mu\nu}$ to yield $f(K,B)$ as in \eqref{eq:f=BF} several of them are related (either exactly, or up to boundary terms). The redundancies are 
\begin{align}\label{eq:f(B)_redundancies}
    B_{\mu\nu} \big( p^1 - p^3 )^{\mu\nu} &= 0 \\
    B_{\mu\nu} \big( q^3 - q^6 )^{\mu\nu} &= 0 \\
    B_{\mu\nu} \big( q^2 + q^5 )^{\mu\nu} &= 0 \\
    B_{\mu\nu} \big( p^2 )^{\mu\nu} &= 0 \\
    B_{\mu\nu} \big( g^1 + g^{10} + p^3 + p^5 - q^1 )^{\mu\nu} &\doteq 0 \\
    B_{\mu\nu} \big( g^2 + g^9 - p^2 - p^4 + p^5 + q^2 )^{\mu\nu} &\doteq 0 \\
    B_{\mu\nu} \big( g^3 + g^8 + p^3 - 2p^5 + q^3 )^{\mu\nu} &\doteq 0 \\
    B_{\mu\nu} \big( 2g^4 + (d-1) p^4 + q^4 )^{\mu\nu} &\doteq 0 \\
    B_{\mu\nu} \big( g^5 - g^6 + (d-1)p^5 - q^6 )^{\mu\nu} &\doteq 0 
\end{align}
where $\doteq$ is an equality up to boundary terms. Since $f(K,B)$ is integrated over the whole spacetime in the action and we assume sufficiently fast vanishing of all fields at infinity, we will neglect such boundary terms.
There are 9 redundancies between the 21 two-forms which can be used to build $F(K,B)$, so we can choose a minimal set of 12 two-forms from which to build it. We choose to parameterise $F(K,B)$ by
\begin{equation}
\begin{split}\label{eq:F_parameterisation}
    F(B)_{\mu\nu} &= d_1 g^1_{\mu\nu} + d_2 g^2_{\mu\nu} + d_3 g^3_{\mu\nu} + d_5 g^5_{\mu\nu} + d_7 g^7_{\mu\nu} \\
    & \quad + s_1 p^1_{\mu\nu} + s_4 p^4_{\mu\nu} + s_5 p^5_{\mu\nu} + r_1 q^1_{\mu\nu} + r_2 q^2_{\mu\nu} + r_3 q^3_{\mu\nu} + r_4 q^4_{\mu\nu}
\end{split}
\end{equation}
where the $d_i$, $s_i$ and $r_i$ are constants.

\subsection{Variation of the general counterterm}

We now study the variation of the general form of $f(K,B)$ found above and demand that it cancels against \eqref{eq:BdY_variation_original}. The variation of $f(K,B)$ is
\begin{equation}\label{eq:f_variation_expand}
    \delta f(K,B) = \delta B_{\mu\nu} F(K,B)^{\mu\nu} + B_{\mu\nu} \delta F(K,B)^{\mu\nu}
\end{equation}
to first order in $\lambda$. In order to evaluate this explicitly, we must study the variations of the basis of 2-forms from which $F(K,B)$ is built. These are
\begin{align}
    B^{\mu\nu} \delta g^1_{\mu\nu} &= \delta B_{\mu\nu} \big( -g^{10} - p^3 - p^5 + q^1 \big)^{\mu\nu} \label{eq:basis_variation_first} \\
    B^{\mu\nu} \delta g^2_{\mu\nu} &= \delta B_{\mu\nu} \big( -g^9 - p^2 + p^4 + p^5 + q^5 \big)^{\mu\nu} \\
    B^{\mu\nu} \delta g^3_{\mu\nu} &= \delta B_{\mu\nu} \big( -g^8 - p^1 +2p^5 - q^6 \big)^{\mu\nu} \\
    B^{\mu\nu} \delta g^4_{\mu\nu} &= \delta B_{\mu\nu} \big( -g^4 - (d-1) p^4 - q^4 \big)^{\mu\nu} \\
    B^{\mu\nu} \delta g^5_{\mu\nu} &= \delta B_{\mu\nu} \big( g^6 - (d-1) p^5 - q^2 \big)^{\mu\nu} \\
    B^{\mu\nu} \delta g^6_{\mu\nu} &= \delta B_{\mu\nu} \big( g^5 + (d-1) p^5 - q^5 \big)^{\mu\nu} \\
    B^{\mu\nu} \delta g^7_{\mu\nu} &= \delta B_{\mu\nu} \big( g^7 + p^2 \big)^{\mu\nu} \\
    B^{\mu\nu} \delta g^8_{\mu\nu} &= \delta B_{\mu\nu} \big( -g^3 - p^3 +2p^5 - q^3 \big)^{\mu\nu} \\
    B^{\mu\nu} \delta g^9_{\mu\nu} &= \delta B_{\mu\nu} \big( -g^2 + p^2 + p^4 + p^5 \big)^{\mu\nu} \\
    B^{\mu\nu} \delta g^{10}_{\mu\nu} &= \delta B_{\mu\nu} \big( -g^1 - p^1 - p^5 + q^1 \big)^{\mu\nu} \\
    B^{\mu\nu} \delta p^1_{\mu\nu} &= \delta B_{\mu\nu} \big( p^3 \big)^{\mu\nu} \\
    B^{\mu\nu} \delta p^2_{\mu\nu} &= \delta B_{\mu\nu} \big( -p^2 \big)^{\mu\nu} \\
    B^{\mu\nu} \delta p^3_{\mu\nu} &= \delta B_{\mu\nu} \big( p^1 \big)^{\mu\nu} \\
    B^{\mu\nu} \delta p^4_{\mu\nu} &= \delta B_{\mu\nu} \big( p^4 \big)^{\mu\nu} \\
    B^{\mu\nu} \delta p^5_{\mu\nu} &= \delta B_{\mu\nu} \big( p^5 \big)^{\mu\nu} \\
    B^{\mu\nu} \delta q^1_{\mu\nu} &= \delta B_{\mu\nu} \big( q^1 \big)^{\mu\nu} \\
    B^{\mu\nu} \delta q^2_{\mu\nu} &= \delta B_{\mu\nu} \big( -q^5 \big)^{\mu\nu} \\
    B^{\mu\nu} \delta q^3_{\mu\nu} &= \delta B_{\mu\nu} \big( q^6 \big)^{\mu\nu} \\
    B^{\mu\nu} \delta q^4_{\mu\nu} &= \delta B_{\mu\nu} \big( q^4 \big)^{\mu\nu} \\
    B^{\mu\nu} \delta q^5_{\mu\nu} &= \delta B_{\mu\nu} \big( -q^2 \big)^{\mu\nu} \\
    B^{\mu\nu} \delta q^6_{\mu\nu} &= \delta B_{\mu\nu} \big( q^3 \big)^{\mu\nu} \label{eq:basis_variation_last}
\end{align}
Inserting these into \eqref{eq:f_variation_expand} yields
\begin{equation}
\begin{split}\label{eq:delta_f_original}
    \delta f(K,B) = \delta B_{\mu\nu} & \big( d_1 g^1 + d_2 g^2 + d_3 g^3 + d_5 g^5 + d_5 g^6 + 2d_7 g^7 - d_3 g^8 - d_2 g^9 - d_1 g^{10} \\
    & + (s_1 - d_3) p^1 + (d_7 - d_2) p^2 + (s_1 - d_1) p^3 + (2s_4 + d_2) p^4 \\
    & + (2s_5 - d_1 + d_2 + 2d_3 -(d-1)d_5 ) p^5 + (2r_1 + d_1) q^1 + (r_2 - d_5) q^2 \\
    & + r_3 q^3 + 2r_4 q^4 + (d_2 - r_2) q^5 + (r_3 - d_3)q^6 \big)^{\mu\nu}
\end{split}
\end{equation}
There are, however, several redundancies which can be identified between the variations. That is, up to boundary terms, we find
\begin{align}
    \delta B_{\mu\nu} \big( 2g^2 + g^4 - 2p^2 + (d-3) p^4 - 2p^5 + 2q^2 + q^4 \big)^{\mu\nu} &\doteq 0 \label{eq:df_redundancy_1} \\
    \delta B_{\mu\nu} \big( g^3 - 2g^6 + p^3 + 2(d-2) p^5 + 2q^2 + q^3 \big)^{\mu\nu} &\doteq 0 \label{eq:df_redundancy_2} \\
    \delta B_{\mu\nu} \big( g^1 - g^5 + g^7 + p^1 + p^2 - (d-2)p^5 - q^1 + q^5 \big)^{\mu\nu} &\doteq 0 \label{eq:df_redundancy_3}
\end{align}
We can use these relations to remove the $g^2$, $q^3$ and $g^7$ contributions to eq.~\eqref{eq:delta_f_original}. The result is 
\begin{align}
\begin{split}
    \delta f(K,B) = \delta B_{\mu\nu} &\big( (d_1 - 2d_7) g^1 + (d_3 - r_3) g^3 - \frac{1}{2} d_2 g^4 + (d_5 + 2d_7) g^5 + (d_5 + 2r_3) g^6 \\
    & - d_3 g^8 - d_2 g^9 - d_1 g^{10} + (s_1 -d_3 - 2d_7) p^1 -d_7 p^2 + (s_1 -d_1 -r_3) p^3 \\
    & + (2s_4 - \frac{1}{2}(d-5) d_2) p^4 + \big[ 2s_5 - d_1 + 2d_2 + 2d_3 - (d-1) d_5 \\
    & + 2(d-2) d_7 - 2(d-2) r_3 \big] p^5 + (2r_1 + d_1 + 2d_7) q^1 + (r_2 -d_2 -d_5 - 2r_3) q^2 \\
    & + (2r_4 - \frac{1}{2} d_2) q^4 + (d_2 -r_2 -2d_7) q^5 + (r_3 - d_3)q^6 \big)^{\mu\nu}
\end{split}
\end{align}
This is the minimal form of the variation of $f(K,B)$ (in the sense that it involves the fewest possible distinct 2-forms $g^i$, $p^i$, $q^i$) and we must impose that it precisely cancels the variation in \eqref{eq:BdY_variation_original}.

\subsection{Fixing \texorpdfstring{$f(K,B)$}{f(K,B)}}

In order to solve for the coefficients $d_i$, $s_i$, $r_i$, we can expand \eqref{eq:BdY_variation_original} in terms of the basis of 2-forms $g^i$, $p^i$, $q^i$. We find that it can be written as
\begin{align}
\begin{split}
    -B_{\mu\nu} \delta Y_+(K)^{\mu\nu} \doteq -2(d-3)\delta B_{\mu\nu} & \big( 2g^1 + g^3 + g^4 + 2g^5 + 2g^8 + 2g^9 + 2g^{10} \\
    & + 4p^1 + 2p^2 + 3p^3 + (d-3) p^4 + 2(d-3) p^5 \\
    & + 2q^1 + 2q^2 + 2q^5 - q^6 \big)^{\mu\nu}
\end{split}
\end{align}
Now, imposing that the variation of $f(K,B)$ cancels this, i.e.,
\begin{equation}
    -B_{\mu\nu}\delta Y_+(K)^{\mu\nu} = -\delta f(K,B)
\end{equation}
gives a system of equations for the coefficients $d_i$, $s_i$, $r_i$. We find that there is a unique solution:
\begin{gather}
    d'_1 = 2 \qc
    d'_2 = 2 \qc 
    d'_3 = 2 \qc
    d'_5 = -6 \qc
    d'_7 = 2 , \nonumber \\
    s'_1 = 2 \qc
    s'_4 = -1 \qc
    s'_5 = 1-3d , \\
    r'_1 = -4 \qc
    r'_2 = 0 \qc
    r'_3 = 3 \qc
    r'_4 = \frac{1}{2} \nonumber
\end{gather}
where $d'_i = -\frac{1}{2(d-3)}d_i$, $s'_i = -\frac{1}{2(d-3)}s_i$, and $r'_i = -\frac{1}{2(d-3)}r_i$.

Substituting these values back into \eqref{eq:F_parameterisation} gives the requisite form of $F(K,B)$ and, from \eqref{eq:f=BF}, of $f(K,B)$. The result can be written
\begin{equation}
\begin{split}\label{eq:f(B)}
    f(K,B) = -B^{\mu\nu} & \big( 2 \partial_\alpha B^{\alpha\beta} K_{\beta[\mu} k_{\nu]} + 2 \partial^\alpha B_{\alpha[\mu} K_{\nu]\beta} k^\beta + 2 \partial^\alpha B_{\alpha\beta} K_{\mu\nu} k^\beta \\
    & -6 \partial_\alpha B_{\beta[\mu} K^{\alpha\beta} k_{\nu]} + 2\partial_\alpha B_{\beta[\mu} K\indices{_{\nu]}^\beta} k^\alpha \\
    & + 2 B^{\alpha\beta} \tilde{K}_{\alpha\beta[\mu} k_{\nu]} - B_{\mu\nu} \hat{K}_\alpha k^\alpha - (3d-1) B_{\alpha[\mu} \hat{K}_{\nu]} k^\alpha \\
    & -4 B_{\alpha\beta} K\indices{^\beta_{[\mu}} \partial\indices{_{\nu]}} k^\alpha + 3B_{\alpha\beta} K_{\mu\nu} \partial^\alpha k^\beta + \frac{1}{2} B_{\mu\nu} K_{\alpha\beta} \partial^\alpha k^\beta \big)
\end{split}
\end{equation}
where we have used \eqref{eq:k=Khat}.
\section{Rewriting the gauged action for two electric 1-form symmetries}
\label{app:gauging_multiple_electric_symmetries}

In this appendix we present the details of finding the action \eqref{eq:S1''_final_form} which is simultaneously gauge-invariant under two electric 1-form symmetries corresponding to Killing vectors $k_I$ and $k_J$. That is, we seek an action which is simultaneously invariant under\footnote{As in the main text, repeated $I$, $J$ indices are \emph{not} summed over unless indicated explicitly.}
\begin{equation}\label{eq:k_transf}
    \delta h_{\mu\nu} = \frac{1}{2} \left( \lambda^I_{\mu} k_{I\nu} + \lambda^I_{\nu} k_{I\mu} \right) \qc \delta B^I_{\mu\nu} = \partial_{[\mu} \lambda^I_{\nu]}
\end{equation}
and
\begin{equation}\label{eq:k'_transf}
    \delta h_{\mu\nu} = \frac{1}{2} \left( \lambda^J_{\mu} k_{J\nu} + \lambda^J_\nu k_{J\mu} \right) \qc \delta B^J_{\mu\nu} = \partial_{[\mu} \lambda^J_{\nu]}
\end{equation}
for unconstrained 1-form parameters $\lambda^I$ and $\lambda^J$. 

Our starting point is the action $S_1$ in \eqref{eq:naive_action}, which we repeat here for convenience,
\begin{equation}\label{eq:naive_action_appendix}
    S_1 = \int\dd[d]{x} \left( \frac{1}{2} h_{\mu\nu} G^{\mu\nu} - B^I_{\mu\nu} Y_+(K_I)^{\mu\nu} + f(K_I,B^I) - B^J_{\mu\nu} Y_+(K_J)^{\mu\nu} + f(K_J,B^J) \right) 
\end{equation}
where $K_I$ and $K_J$ are $\B$- or $\D$-type CKY tensors related to $k_I$ and $k_J$ respectively by \eqref{eq:k=Khat}.
As mentioned in section~\ref{sec:gauging_multiple_symms}, this is not invariant under either symmetry \eqref{eq:k_transf} and \eqref{eq:k'_transf}.
We will proceed by first finding counter-terms which render the action invariant under the \eqref{eq:k_transf} transformations. We will then find counter-terms which will make the action invariant under the \eqref{eq:k'_transf} transformations, without spoiling invariance under \eqref{eq:k_transf}.

The variation of $S_1$ under the \eqref{eq:k_transf} transformation comes from the $B^J_{\mu\nu} Y_+(K_J)^{\mu\nu}$ term. 
Integrating by parts and using the fact that $k_I$ is a Killing vector, we find that the variation can be written
\begin{equation}
\begin{split}\label{eq:B'dY'_variation}
    -\int \dd[d]{x} B^J_{\mu\nu} \delta Y_+(K_J)^{\mu\nu} &= -6 \int \dd[d]{x} \partial_{[\alpha} \lambda^I_{\beta]} \Big( \partial^{[\alpha} B^J_{\mu\nu} K_J^{\mu\nu} k_I^{\beta]} + B^J_{\mu\nu} \tilde{K}_J^{[\alpha\mu\nu} k_I^{\beta]} \\
    &\quad + \partial^{[\alpha} B^{J\mu\nu} K_{J\mu\nu} k_I^{\beta]} + B^{J[\mu\nu} \partial^\alpha K_{J\mu\nu} k_I^{\beta]} + 2\eta^{\beta[\alpha} B^{J\rho\sigma} K_{J\rho\sigma} \partial^{\delta]} k_{I\delta} \Big)
\end{split}
\end{equation}
The manipulations leading to this result are completely analogous to those in \eqref{eq:BdY_variation_original}. Hence, a counter-term action which cancels this variation is
\begin{equation}
\begin{split}\label{eq:S_ct}
    S_{\text{c.t.}} &= 6 \int \dd[d]{x} B^I_{\alpha\beta} \Big( \partial^{[\alpha} B^J_{\mu\nu} K_J^{\mu\nu} k_I^{\beta]} + B^J_{\mu\nu} \tilde{K}_J^{[\alpha\mu\nu} k_I^{\beta]} \\
    &\quad + \partial^{[\alpha} B^{J\mu\nu} K_{J\mu\nu} k_I^{\beta]} + B^{J[\mu\nu} \partial^\alpha K_{J\mu\nu} k_I^{\beta]} + 2\eta^{\beta[\alpha} B^{J\rho\sigma} K_{J\rho\sigma} \partial^{\delta]} k_{I\delta} \Big)
\end{split}
\end{equation}
Therefore, the action
\begin{equation}\label{eq:S_K_gauged}
    S_1' = S_1 + S_{\text{c.t.}}
\end{equation}
is invariant under \eqref{eq:k_transf}. We must now study the variation of $S_1'$ under \eqref{eq:k'_transf}. 
There are two contributions: one from the $B^I_{\mu\nu}Y_+(K_I)^{\mu\nu}$ term in $S_1$, and the other from $S_{\text{c.t.}}$. The former is analogous to \eqref{eq:B'dY'_variation}, and can be written
\begin{align}
\begin{split}\label{eq:BdY_variation}
    -\int \dd[d]{x} B^I_{\mu\nu} \delta Y_+(K_I)^{\mu\nu} &= -6 \int \dd[d]{x} \partial_{[\alpha} \lambda^J_{\beta]} \Big( \partial^{[\alpha} B^I_{\mu\nu} K_I^{\mu\nu} k_J^{\beta]} + B^I_{\mu\nu} \tilde{K}_I^{[\alpha\mu\nu} k_J^{\beta]} \\
    &\quad + \partial^{[\alpha} B^{I\mu\nu} K_{I\mu\nu} k_J^{\beta]} + B^{I[\mu\nu} \partial^\alpha K_{I\mu\nu} k_J^{\beta]} + 2\eta^{\beta[\alpha} B^{I\rho\sigma} K_{I\rho\sigma} \partial^{\delta]} k_{J\delta} \Big)
\end{split}
\end{align}
whereas the variation of $S_{\text{c.t.}}$ is
\begin{align}
\begin{split}
    \delta S_{\text{c.t.}} &= -6 \int\dd[d]{x} \partial_{[\alpha} \lambda^J_{\beta]} \Big( \partial^{[\mu} B^I_{\mu\nu} K_J^{\alpha\beta} k_I^{\nu]} + B^I_{\mu\nu} K_J^{[\alpha\beta} \partial^\mu k_I^{\nu]} \\
    &\quad - B^{I[\mu\nu} \partial_\mu K_J^{\alpha\beta]} k_{I\nu} + 2 B\indices{^I_{\mu}^{[\mu}} K_J^{\alpha\beta} \partial^{\delta]}k_{I\delta} \Big)
\end{split}
\end{align}
The sum of these two variations can be written
\begin{equation}
\begin{split}\label{eq:S+Sct_var}
    \delta S_1' &= -6 \int\dd[d]{x} \delta B^J_{\alpha\beta} \delta^{\mu\nu\alpha\beta}_{\rho\sigma\gamma\delta} \bigg[ (\dd{B^I})^{\gamma\rho\sigma} K_{I\mu\nu} k_J^\delta + 3(\dd{B^I})\indices{^\gamma_{\mu\nu}} \left( K_I^{\rho\sigma}k_J^\delta + K_J^{\rho\sigma} k_I^\delta \right) \\
    &\quad + 2B\indices{^I_\nu^\gamma} \tilde{K}\indices{_{I\mu}^{\rho\sigma}} k_J^\delta + 2B\indices{^I_\nu^\gamma} \tilde{K}\indices{_{J\mu}^{\rho\sigma}} k_I^\delta + 2B\indices{^I_\nu^\gamma} K_I^{\rho\sigma} \partial_\mu k_J^\delta + B^I_{\mu\nu} \tilde{K}_I^{\gamma\rho\sigma} k_J^\delta + B^{I\rho\sigma} \tilde{K}\indices{_I^\gamma_{\mu\nu}} k_J^\delta \\
    &\quad - B^{I\rho\sigma}\tilde{K}\indices{_{J\mu}^{\gamma\delta}} k_{I\nu} + B^{I\rho\sigma} K_{I\mu\nu} \partial^\delta k_J^\gamma - 2B^{I\rho\sigma} K\indices{_{I\mu}^\gamma} \partial^\delta k_{J\nu} + B^I_{\mu\nu} K_J^{\rho\sigma} \partial^\gamma k_I^\delta \bigg]
\end{split}
\end{equation}
where several terms have been integrated by parts and rearranged using the property
\begin{equation}\label{eq:eta_property}
    \eta^{[\mu\nu\alpha\beta|\rho]\sigma\gamma\delta} = 0
\end{equation}
of the $\eta$ symbol, which is related to the generalised Kronecker delta $\delta^{\mu\nu\alpha\beta}_{\rho\sigma\gamma\delta}$ by raising/lowering indices. 
In particular, \eqref{eq:eta_property} implies that
\begin{equation}\label{eq:eta_property2}
    3 \eta^{\mu\nu[\alpha\beta|\gamma]\delta\rho\sigma} + 2 \eta^{\alpha\beta\gamma[\mu|\nu]\delta\rho\sigma} = 0
\end{equation}
which is used regularly to rearrange terms in this appendix.

The first two terms in square parentheses in \eqref{eq:S+Sct_var} depend only on $\dd{B^I}$ and so can be cancelled by the addition of a counter-term
\begin{equation}\label{eq:Sct'}
    S_{\text{c.t.}}' = 6 \int \dd[d]{x} B^J_{\alpha\beta} \delta^{\mu\nu\alpha\beta}_{\rho\sigma\gamma\delta} \bigg[ (\dd{B^I})^{\gamma\rho\sigma} K_{I\mu\nu} k_J^\delta + 3(\dd{B^I})\indices{^\gamma_{\mu\nu}} \left( K_I^{\rho\sigma} k_J^\delta + K_J^{\rho\sigma} k_I^\delta \right) \bigg]
\end{equation}
which, since it depends only on $\dd{B^I}$, is invariant under \eqref{eq:k_transf}.
So the action 
\begin{equation}\label{eq:S1''}
    S_1'' = S_1 + S_{\text{c.t.}} + S_{\text{c.t.}}'
\end{equation}
remains invariant under \eqref{eq:k_transf}, and under \eqref{eq:k'_transf} it transforms as
\begin{equation}
\begin{split}\label{eq:S1''_var}
    \delta S_1'' &= -6 \int\dd[d]{x} \delta B^J_{\alpha\beta} \delta^{\mu\nu\alpha\beta}_{\rho\sigma\gamma\delta} \bigg[ 2B\indices{^I_\nu^\gamma} \tilde{K}\indices{_{I\mu}^{\rho\sigma}} k_J^\delta + 2B\indices{^I_\nu^\gamma} \tilde{K}\indices{_{J\mu}^{\rho\sigma}} k_I^\delta +2B\indices{^I_\nu^\gamma} K_I^{\rho\sigma} \partial_\mu k_J^\delta \\
    &\qquad + B^I_{\mu\nu} \tilde{K}_I^{\gamma\rho\sigma} k_J^\delta + B^{I\rho\sigma} \tilde{K}\indices{_I^\gamma_{\mu\nu}} k_J^\delta - B^{I\rho\sigma} \tilde{K}\indices{_{J\mu}^{\gamma\delta}} k_{I\nu} + B^{I\rho\sigma} K_{I\mu\nu} \partial^\delta k_J^\gamma \\
    &\qquad - 2B^{I\rho\sigma} K\indices{_{I\mu}^\gamma} \partial^\delta k_{J\nu} + B^I_{\mu\nu} K_J^{\rho\sigma} \partial^\gamma k_I^\delta \bigg]
\end{split}
\end{equation}
We will now show, by using relations between the Minkowski space CKY tensors, that this variation vanishes for all $k_I$ and $k_J$. This relies on the specific form of the solutions \eqref{eq:CKY_solution} to the CKY equation \eqref{eq:CKY_equation} on Minkowski space. Therefore, it is simplest to show the vanishing of \eqref{eq:S1''_var} case-by-case.

\paragraph{Case 1: $k_I$ and $k_J$ constant.}
When both $k_I$ and $k_J$ are constant Killing vectors, the analysis is straightforward. In this case, both $K_I$ and $K_J$ are $\B$-type CKY tensors (see \eqref{eq:Khat_solution}), for which $\tilde{K}_I=\tilde{K}_J=0$. This immediately implies that the variation \eqref{eq:S1''_var} vanishes. In fact, there is another simplification in this case as $K_I^{[\rho\sigma} k_J^{\delta]} + K_J^{[\rho\sigma} k_I^{\delta]} = 0$, so the second term in square parentheses in \eqref{eq:Sct'} vanishes, so the action $S_1''$ takes a slightly simpler form.

\paragraph{Case 2: $k_I$ not constant and $k_J$ constant.}
Next consider the case where $k_J$ is a constant Killing vector but $k_I$ is not. In this case, $K_J$ is a $\B$-type CKY tensor while $K_I$ is $\D$-type, explicitly we parameterise them by
\begin{equation}
\begin{split}\label{eq:KK'_k'const}
    K_{I\mu\nu} &= 2x_{[\mu} \D_{I\nu]\lambda}x^\lambda + \frac{1}{2} \D_{I\mu\nu} x^2 \\
    K_{J\mu\nu} &= \B_{J[\mu}x_{\nu]}
\end{split}
\end{equation}
where $\D_I$ is a constant 2-form and $\B_J$ is a constant vector (see \eqref{eq:CKY_solution}). Note that $\tilde{K}_J = 0$ in this case.
Several terms in the variation \eqref{eq:S1''_var} then vanish, and those that remain can be rearranged using $\eta^{[\mu\nu\alpha\beta|\rho]\sigma\gamma\delta}=0$ to give
\begin{equation}\label{eq:S1''_var_k'const}
    \delta S_1'' = -6 \int \dd[d]{x} \delta B^J_{\alpha\beta} \delta^{\mu\nu\alpha\beta}_{\rho\sigma\gamma\delta} B^I_{\mu\nu} \left( K_J^{\rho\sigma} \partial^\gamma k_I^\delta + \frac{2}{3} \tilde{K}_I^{\rho\sigma\gamma} k_J^\delta \right)
\end{equation}
From \eqref{eq:KK'_k'const}, we find the relations
\begin{equation}
    \tilde{K}_{I\mu\nu\rho} = 3\D_{I[\mu\nu}x_{\rho]}\qc \partial_\mu k_{I\nu} = -2(d-3)\D_{I\mu\nu} \qc k_{J\mu} = -(d-3) \B_{J\mu}
\end{equation}
Substituting these into \eqref{eq:S1''_var_k'const} then gives a vanishing result.

\paragraph{Case 3: $k_I$ constant and $k_J$ not constant.}
We must find the same result with $k_I$ and $k_J$ interchanged, so we should find that the variation vanishes also when $k_I$ is a constant Killing vector but $k_J$ is not. In this case, $K_I$ is a $\B$-type CKY tensor and $K_J$ is $\D$-type, parameterised by
\begin{equation}
\begin{split}\label{eq:KK'_kconst}
    K_{I\mu\nu} &= \B_{I[\mu}x_{\nu]} \\
    K_{J\mu\nu} &= 2x_{[\mu} \D_{J\nu]\lambda}x^\lambda + \frac{1}{2} \D_{J\mu\nu} x^2 
\end{split}
\end{equation}
with $\B_I$ a constant vector and $\D_J$ a constant 2-form. Now we have $\tilde{K}_I=0$, so \eqref{eq:S1''_var} reduces to
\begin{equation}
\begin{split}\label{eq:S1''_var_kconst}
    \delta S_1'' &= -6 \int\dd[d]{x} \delta B^J_{\alpha\beta} \delta^{\mu\nu\alpha\beta}_{\rho\sigma\gamma\delta} B^{I\rho\sigma} \big( -2 \tilde{K}\indices{_{J\mu}^{\gamma\delta}} k_{I\nu} - 2 \tilde{K}\indices{_{J\mu\nu}^\gamma} k_I^\delta - K_{I\mu\nu} \partial^\gamma k_J^\delta \\
    &\qquad\qquad - K_I^{\gamma\delta} \partial_\mu k_{J\nu} + 4 K\indices{_{I\mu}^\gamma} \partial_\nu k_J^\delta \big)
\end{split}
\end{equation}
From \eqref{eq:KK'_kconst}, we now have the relations
\begin{equation}\label{eq:KK'_kconst_relations}
    \tilde{K}_{J\mu\nu\rho} = 3\D_{J[\mu\nu}x_{\rho]}\qc \partial_\mu k_{J\nu} = -2(d-3)\D_{J\mu\nu}\qc k_{I\mu} = -(d-3) \B_{I\mu}
\end{equation}
Substituting these into \eqref{eq:S1''_var_kconst} again gives a vanishing result.

\paragraph{Case 4: $k_I$ and $k_J$ not constant.}
Finally, we consider the case where both $k_I$ and $k_J$ are non-constant Killing vectors, so both $K_I$ and $K_J$ are $\D$-type CKY tensors, parameterised by two constant 2-forms $\D_I$ and $\D_J$ as
\begin{equation}
\begin{split}\label{eq:KK'_nonconst}
    K_{I\mu\nu} &= 2x_{[\mu} \D_{I\nu]\lambda}x^\lambda + \frac{1}{2} \D_{I\mu\nu} x^2 \\
    K_{J\mu\nu} &= 2x_{[\mu} \D_{J\nu]\lambda}x^\lambda + \frac{1}{2} \D_{J\mu\nu} x^2
\end{split}
\end{equation}
In this case the vanishing of the variation \eqref{eq:S1''_var} is harder to see. Rearranging in a similar manner to \eqref{eq:S1''_var_k'const} and \eqref{eq:S1''_var_kconst}, the variation \eqref{eq:S1''_var} can be written
\begin{equation}
\begin{split}\label{eq:S1''_var_nonconst}
    \delta S_1'' &= -6 \int \dd[d]{x} \delta B^J_{\alpha\beta} \delta^{\mu\nu\alpha\beta}_{\rho\sigma\gamma\delta} \bigg[ B^I_{\mu\nu} \left( K_J^{\rho\sigma} \partial^\gamma k_I^\delta + \frac{2}{3} \tilde{K}_I^{\rho\sigma\gamma} k_J^\delta \right) \\
    &\quad + B^{I\rho\sigma} \left( -2 \tilde{K}\indices{_{J\mu}^{\gamma\delta}} k_{I\nu} - 2 \tilde{K}\indices{_{J\mu\nu}^\gamma} k_I^\delta - K_{I\mu\nu} \partial^\gamma k_J^\delta - K_I^{\gamma\delta} \partial_\mu k_{J\nu} + 4 K\indices{_{I\mu}^\gamma} \partial_\nu k_J^\delta \right) \bigg]
\end{split}
\end{equation}
From \eqref{eq:KK'_nonconst}, we have
\begin{equation}
    k_{I\mu} = 2(d-3) \D_{I\mu\nu}x^\nu\qc \tilde{K}_{I\mu\nu\rho} = 3\D_{I[\mu\nu}x_{\rho]}
\end{equation}
and similarly for $K_J$.
Substituting these relations into \eqref{eq:S1''_var_nonconst}, we find
\begin{equation}\label{eq:S1''_var_Dtype}
    \delta S_1'' = -6 (d-3) \int\dd[d]{x} \delta B^J_{\alpha\beta} \delta^{\mu\nu\alpha\beta}_{\rho\sigma\gamma\delta} \bigg[ -B^I_{\mu\nu} (\D_I\wedge \D_J)^{\rho\sigma\gamma\delta} + 6 B^{I\rho\sigma} (\D_I\wedge \D_J)\indices{_{\mu\nu}^{\gamma\delta}} \bigg] x^2
\end{equation}
where $(\D_I\wedge\D_J)_{\mu\nu\rho\sigma} = \D_{I[\mu\nu} \D_{J\rho\sigma]}$.
Now, it follows from $\eta^{\alpha\beta\mu[\nu|\gamma\delta\rho\sigma]}=0$ that
\begin{equation}\label{eq:eta_symmetry}
    6 \eta^{\alpha\beta[\mu\nu|\gamma\delta]\rho\sigma} = \eta^{\alpha\beta\rho\sigma|\mu\nu\gamma\delta}
\end{equation}
Inserting this relation in \eqref{eq:S1''_var_Dtype} then gives a vanishing result.
Therefore, we have shown that for all types of Killing vector $k_I$ and $k_J$, the action $S_1''$ in \eqref{eq:S1''} is invariant under both \eqref{eq:k_transf} and \eqref{eq:k'_transf}. Therefore, there is no mixed 't Hooft anomaly between any pair of 1-form symmetries \eqref{eq:dh=lambda_k}.

\subsection{Rewriting the gauge-invariant action}

The construction of $S_1''$ above treated the two symmetries \eqref{eq:k_transf} and \eqref{eq:k'_transf} separately and as a result the action is not obviously symmetric under exchange of $(k_I,B^I)\leftrightarrow(k_J,B^J)$. 
The action $S_1''$ in \eqref{eq:S1''} is made up of three terms: $S_1$, $S_{\text{c.t.}}$ and $S_{\text{c.t.}}'$, given by \eqref{eq:naive_action_appendix}, \eqref{eq:S_ct} and \eqref{eq:Sct'} respectively. Examining these individually, we see that $S_1$ is symmetric under interchange $(k_I,B^I) \leftrightarrow (k_J,B^J)$, whereas $S_{\text{c.t.}}$ and $S_{\text{c.t.}}'$ are not manifestly so. We will now show that $S_{\text{c.t.}} + S_{\text{c.t.}}'$ can, in fact, be written in the more symmetric form
\begin{equation}
    S_{\text{c.t.}} + S_{\text{c.t.}}' = s(k_I,B^I,k_J,B^J)
\end{equation}
where
\begin{equation}
\begin{split}\label{eq:s(k,k')_def}
    s(k_I,B^I,k_J,B^J) &= \frac{3}{2} \int \dd[d]{x} \delta_{\mu\nu\rho}^{\alpha\beta\gamma} \left( B^I_{\alpha\beta} B^{J\mu\nu} k_{J\gamma} k_I^\rho - 2B^I_{\alpha\beta} B\indices{^J_\gamma^\mu} k_J^\nu k_I^\rho \right) \\
    &\quad + 6 \int \dd[d]{x} \delta^{\mu\nu\alpha\beta}_{\rho\sigma\gamma\delta} \left( 3(\dd{B^J})\indices{^\gamma_{\mu\nu}} B^I_{\alpha\beta} K_J^{\rho\sigma} k_I^\delta + 3(\dd{B^I})\indices{^\gamma_{\mu\nu}} B^J_{\alpha\beta} K_I^{\rho\sigma} k_J^\delta \right) \\
    &\quad + 6 \int \dd[d]{x} \delta^{\mu\nu\alpha\beta}_{\rho\sigma\gamma\delta} \left( B^I_{\alpha\beta} (\dd{B^J})^{\gamma\rho\sigma} K_{J\mu\nu} k_I^\delta + B^J_{\alpha\beta} (\dd{B^I})^{\gamma\rho\sigma} K_{I\mu\nu} k_J^\delta \right) \\
    &\quad + 6 \int \dd[d]{x} \delta^{\mu\nu\alpha\beta}_{\rho\sigma\gamma\delta} B^I_{\alpha\beta} B^J_{\mu\nu} \left( \frac{2}{3} \tilde{K}_J^{\rho\sigma\gamma} k_I^\delta - K_J^{\rho\sigma} \partial^\gamma k_I^\delta \right) 
\end{split}
\end{equation}
such that, from \eqref{eq:naive_action},
\begin{equation}\label{eq:S1''_nice}
    S_1'' = S_0 + s(k_I,B^I) + s(k_J,B^J) + s(k_I,B^I,k_J,B^J)
\end{equation}
Then, in the following subsection, we will show that $s(k_I,B^I,k_J,B^J)$ is invariant under exchange of $(k_I,B^I)$ with $(k_J,B^J)$.

Consider first $S_{\text{c.t.}}$ defined in \eqref{eq:S_ct}. This can be more usefully written by using the generalised Kronecker delta function as
\begin{equation}
\begin{split}\label{eq:S_ct_working}
    S_{\text{c.t.}} &= 6 \int \dd[d]{x} \delta^{\mu\nu\alpha\beta}_{\rho\sigma\gamma\delta} B^I_{\alpha\beta} \bigg( \partial^\gamma B^J_{\mu\nu} K_J^{\rho\sigma} k_I^\delta + B^J_{\mu\nu} \tilde{K}_J^{\rho\sigma\gamma} k_I^\delta + (\dd{B^J})^{\gamma\rho\sigma} K_{J\mu\nu} k_I^\delta + B^{J\rho\sigma} \tilde{K}\indices{_J^\gamma_{\mu\nu}} k_I^\delta \\
    &\quad - B^{J\rho\sigma} K_{J\mu\nu} \partial^\gamma k_I^\delta + 2 B^{J\rho\sigma} K\indices{_{J\mu}^\gamma} \partial_\nu k_I^\delta \bigg) + \frac{3}{2} \int \dd[d]{x} \delta^{\alpha\beta\gamma}_{\mu\nu\rho} B^I_{\alpha\beta} B^{J\mu\nu} k_{J\gamma} k_I^\rho
\end{split}
\end{equation}
where we have the fact that $K_I$ and $K_J$ are CKY tensors and so satisfy \eqref{eq:CKY_equation}. We have also used \eqref{eq:eta_property} to rearrange terms. 

We rearrange the first term in parentheses in \eqref{eq:S_ct_working} by writing $\partial_\gamma B^J_{\mu\nu} = 3 (\dd{B^J})_{\gamma\mu\nu} - 2\partial_{[\mu} B^J_{\nu]\gamma}$ and integrating by parts. Doing so leads to several cancellations. Combining with $S'_{\text{c.t.}}$ in \eqref{eq:Sct'}, the result is
\begin{equation}
\begin{split}\label{eq:S_ct_working2}
    S_{\text{c.t.}} + S'_{\text{c.t.}} &= \frac{3}{2} \int \dd[d]{x} \delta^{\alpha\beta\gamma}_{\mu\nu\rho} \left( B^I_{\alpha\beta} B^{J\mu\nu} k_{J\gamma} k_I^\rho -2 B^I_{\alpha\beta} B\indices{^J_\gamma^\mu} k_J^\nu k_I^\rho \right) \\
    &\quad + 6 \int\dd[d]{x} \delta^{\mu\nu\alpha\beta}_{\rho\sigma\gamma\delta} \bigg( 3(\dd{B^J})\indices{^\gamma_{\mu\nu}} B^I_{\alpha\beta} K_J^{\rho\sigma} k_I^\delta + 3(\dd{B^I})\indices{^\gamma_{\mu\nu}} B^J_{\alpha\beta} K_I^{\rho\sigma} k_J^\delta \\
    &\quad + B^I_{\alpha\beta} (\dd{B^J})^{\gamma\rho\sigma} K_{J\mu\nu} k_I^\delta + B^J_{\alpha\beta} (\dd{B^I})^{\gamma\rho\sigma} K_{I\mu\nu} k_J^\delta + 2B^I_{\alpha\beta} B\indices{^J_\nu^\gamma} \tilde{K}\indices{_{J\mu}^{\rho\sigma}} k_I^\delta \\
    &\quad + B^I_{\alpha\beta} B^{J\rho\sigma} \tilde{K}\indices{_J^\gamma_{\mu\nu}} k_I^\delta + B^I_{\alpha\beta} B^J_{\mu\nu} \tilde{K}_J^{\rho\sigma\gamma} k_I^\delta + 2B^I_{\alpha\beta} B\indices{^J_\nu^\gamma} K_J^{\rho\sigma}\partial_\mu k_I^\delta \\
    &\quad - B^I_{\alpha\beta} B^{J\rho\sigma} K_{J\mu\nu} \partial^\gamma k_I^\delta + 2 B^I_{\alpha\beta} B^{J\rho\sigma} K\indices{_{J\mu}^\gamma} \partial_\nu k_I^\delta \bigg)
\end{split}
\end{equation}
Now, the identity \eqref{eq:eta_property2} implies the relation
\begin{equation}\label{eq:relation1}
    2 \delta^{\mu\nu\alpha\beta}_{\rho\sigma\gamma\delta} B^I_{\alpha\beta} B\indices{^J_\nu^\gamma} \tilde{K}\indices{_{J\mu}^{\rho\sigma}} k_I^\delta = -\delta^{\mu\nu\alpha\beta}_{\rho\sigma\gamma\delta} B^I_{\alpha\beta} B^{J\rho\sigma} \left( \tilde{K}\indices{_{J\mu}^{\gamma\delta}} k_{I\nu} + 2 \tilde{K}\indices{_J^\gamma_{\mu\nu}} k_I^\delta \right)
\end{equation}
as well as
\begin{equation}
\begin{split}\label{eq:relation2}
    2 \delta^{\mu\nu\alpha\beta}_{\rho\sigma\gamma\delta} B^I_{\alpha\beta} B\indices{^J_\nu^\gamma} \tilde{K}\indices{_{J\mu}^{\rho\sigma}} k_I^\delta &= \delta^{\mu\nu\alpha\beta}_{\rho\sigma\gamma\delta} \big( - B^I_{\alpha\beta} B^J_{\mu\nu} \tilde{K}_J^{\rho\sigma\gamma} k_I^\delta + B^{I\gamma\delta} B^J_{\alpha\beta} \tilde{K}\indices{_{J\mu}^{\rho\sigma}} k_{I\nu} \\
    &\qquad + 2 B^{I\gamma\delta} B^J_{\alpha\beta} \tilde{K}\indices{_{J\mu\nu}^\rho} k_I^\sigma \big)
\end{split}
\end{equation}
Equating \eqref{eq:relation1} and \eqref{eq:relation2} gives the identity
\begin{equation}\label{eq:relation4}
    \delta^{\mu\nu\alpha\beta}_{\rho\sigma\gamma\delta} \left( B^I_{\alpha\beta} B^J_{\mu\nu} \tilde{K}_J^{\rho\sigma\gamma} k_I^\delta -3B^I_{\alpha\beta}B^{J\rho\sigma} \left(\tilde{K}\indices{_{J\mu}^{\gamma\delta}} k_{I\nu} + \tilde{K}\indices{_J^\gamma_{\mu\nu}} k_I^\delta \right) \right) = 0
\end{equation}
Furthermore, \eqref{eq:eta_property2} implies
\begin{equation}\label{eq:relation3}
    2 \delta^{\mu\nu\alpha\beta}_{\rho\sigma\gamma\delta} B^I_{\alpha\beta} B\indices{^J_\nu^\gamma} K_J^{\rho\sigma} \partial_\mu k_I^\delta = \delta^{\mu\nu\alpha\beta}_{\rho\sigma\gamma\delta} B^I_{\alpha\beta} B^{J\rho\sigma} \left( - K_J^{\gamma\delta} \partial_\mu k_{I\nu} + 2 K\indices{_{J\mu}^\gamma} \partial_\nu k_I^\delta \right)
\end{equation}
Now, substituting \eqref{eq:relation1} and \eqref{eq:relation3} into \eqref{eq:S_ct_working2} gives
\begin{equation}
\begin{split}
    S_{\text{c.t.}} + S'_{\text{c.t.}} &= \frac{3}{2} \int \dd[d]{x} \delta^{\alpha\beta\gamma}_{\mu\nu\rho} \left( B^I_{\alpha\beta} B^{J\mu\nu} k_{J\gamma} k_I^\rho -2 B^I_{\alpha\beta} B\indices{^J_\gamma^\mu} k_J^\nu k_I^\rho \right) \\
    &\quad + 6 \int\dd[d]{x} \delta^{\mu\nu\alpha\beta}_{\rho\sigma\gamma\delta} \bigg( 3(\dd{B^J})\indices{^\gamma_{\mu\nu}} B^I_{\alpha\beta} K_J^{\rho\sigma} k_I^\delta + 3(\dd{B^I})\indices{^\gamma_{\mu\nu}} B^J_{\alpha\beta} K_I^{\rho\sigma} k_J^\delta \\
    &\quad + B^I_{\alpha\beta} (\dd{B^J})^{\gamma\rho\sigma} K_{I\mu\nu} k_I^\delta + B^J_{\alpha\beta} (\dd{B^I})^{\gamma\rho\sigma} K_{I\mu\nu} k_J^\delta \\
    &\quad - B^I_{\alpha\beta} B^{J\rho\sigma} \left( \tilde{K}\indices{_{J\mu}^{\gamma\delta}} k_{I\nu} + \tilde{K}\indices{_J^\gamma_{\mu\nu}} k_I^\delta \right) + B^I_{\alpha\beta} B^J_{\mu\nu} \tilde{K}_J^{\rho\sigma\gamma} k_I^\delta \\
    &\quad - 6 B^I_{\alpha\beta} B^{J\rho\sigma} (K_J \wedge \dd{k_I})\indices{_{\mu\nu}^{\gamma\delta}} \bigg)
\end{split}
\end{equation}
Finally, using \eqref{eq:relation4} and \eqref{eq:eta_symmetry} gives the result
\begin{equation}
    S_{\text{c.t.}} + S'_{\text{c.t.}} = s(k_I,B^I,k_J,B^J)
\end{equation}
where $s(k_I,B^I,k_J,B^J)$ is given in \eqref{eq:s(k,k')_def}.

\subsection{Exchange symmetry of the gauged action}
\label{app:exchange_symmetry}

We now consider the properties of the gauged action $S_1''$ in \eqref{eq:S1''_nice} under exchange
\begin{equation}\label{eq:interchange}
    (k_I,B^I) \leftrightarrow (k_J,B^J)
\end{equation}
The action $S_1''$ will be symmetric under this exchange if and only if $s(k_I,B^I,k_J,B^J)$ is symmetric, since both $S_0$ and $s(k_I,B^I) + s(k_J,B^J)$ are symmetric individually.
Let us consider how each term of $s(k_I,B^I,k_J,B^J)$ in \eqref{eq:s(k,k')_def} transforms under the exchange \eqref{eq:interchange}.
The two terms in the first integral in \eqref{eq:s(k,k')_def} are both invariant under \eqref{eq:interchange}. This can be seen straightforwardly from the symmetries of $\delta^{\alpha\beta\gamma}_{\mu\nu\rho}$ and the fact that $\eta_{[\mu\nu\rho|\alpha]\beta\gamma}=0$. The second integral in \eqref{eq:s(k,k')_def} is also invariant under \eqref{eq:interchange} as the two terms in the integrand transform into each other. The same is true of the third integral. This leaves the final term in \eqref{eq:s(k,k')_def},
\begin{equation}\label{eq:last_term}
    \delta^{\mu\nu\alpha\beta}_{\rho\sigma\gamma\delta} B^I_{\alpha\beta} B^J_{\rho\sigma} \left( \frac{2}{3} \tilde{K}_J^{\rho\sigma\gamma} k_I^\delta - K_J^{\rho\sigma} \partial^\gamma k_I^\delta \right)
\end{equation}
which is not manifestly symmetric under \eqref{eq:interchange}. However, we can verify that for each possible choice of $k_I$ and $k_J$ that it is invariant under interchange. There are three independent cases to check.

\paragraph{Case 1: $k_I$ and $k_J$ constant.}
When $k_I$ and $k_J$ are constant, the CKY tensors $K_I$ and $K_J$ which relate to them as in \eqref{eq:k=Khat} are both $\B$-type, and so we have $\tilde{K}_I=\tilde{K}_J=0$. Therefore, \eqref{eq:last_term} vanishes in this simple case, and so is clearly invariant under the interchange \eqref{eq:interchange}.

\paragraph{Case 2: $k_I$ constant and $k_J$ not constant.}

In this case, $k_I$ corresponds to a $\B$-type CKY tensor $K_I$ and $k_J$ corresponds to a $\D$-type CKY tensor $K_J$. They can be parameterised as in \eqref{eq:KK'_kconst} and satisfy the relations in \eqref{eq:KK'_kconst_relations}. In this case, \eqref{eq:last_term} evaluates to
\begin{equation}\label{eq:interchange_case2_working}
    \delta^{\mu\nu\alpha\beta}_{\rho\sigma\gamma\delta} B^I_{\alpha\beta} B^J_{\rho\sigma} \left( \frac{2}{3} \tilde{K}_J^{\rho\sigma\gamma} k_I^\delta - K_J^{\rho\sigma} \partial^\gamma k_I^\delta \right) = -2(d-3) \delta^{\mu\nu\alpha\beta}_{\rho\sigma\gamma\delta} B^I_{\alpha\beta} B^J_{\mu\nu} \D_J^{\rho\sigma} x^\gamma \B_I^\delta 
\end{equation}
After the interchange \eqref{eq:interchange}, this evaluates to
\begin{equation}
    \delta^{\mu\nu\alpha\beta}_{\rho\sigma\gamma\delta} B^J_{\alpha\beta} B^I_{\rho\sigma} \left( \frac{2}{3} \tilde{K}_I^{\rho\sigma\gamma} k_J^\delta - K_I^{\rho\sigma} \partial^\gamma k_J^\delta \right) = -2(d-3) \delta^{\mu\nu\alpha\beta}_{\rho\sigma\gamma\delta} B^I_{\alpha\beta} B^J_{\mu\nu} \D_J^{\rho\sigma} x^\gamma \B_I^\delta 
\end{equation}
which agrees with \eqref{eq:interchange_case2_working}. Therefore, \eqref{eq:last_term} is invariant under \eqref{eq:interchange} in this case.

\paragraph{Case 3: $k_I$ and $k_J$ not constant.}

In this case, both $k_I$ and $k_J$ correspond to $\D$-type CKY tensors $K_I$ and $K_J$ which can be parameterised as in \eqref{eq:KK'_nonconst}. This implies the relations \eqref{eq:S1''_var_nonconst} for $K_I$ and there are analogous relations for $K_J$. Substituting these into \eqref{eq:last_term}, we find
\begin{equation}
    \delta^{\mu\nu\alpha\beta}_{\rho\sigma\gamma\delta} B^I_{\alpha\beta} B^J_{\rho\sigma} \left( \frac{2}{3} \tilde{K}_J^{\rho\sigma\gamma} k_I^\delta - K_J^{\rho\sigma} \partial^\gamma k_I^\delta \right) = (d-3) \delta^{\mu\nu\alpha\beta}_{\rho\sigma\gamma\delta} B^I_{\alpha\beta} B^J_{\rho\sigma} \left( 10 \D_J^{[\rho\sigma} \D_I^{\gamma\delta} x^{\lambda]} x_\lambda - \D_J^{\rho\sigma} \D_I^{\gamma\delta} x^2 \right)
\end{equation}
which is manifestly invariant under exchanging $B^I \leftrightarrow B^J$ and $\D_I \leftrightarrow \D_J$ (which is equivalent to $k_I\leftrightarrow k_J$ in this case). Therefore, \eqref{eq:last_term} is again symmetric under \eqref{eq:interchange} in this case.

In summary, $s(k_I,B^I,k_J,B^J)$ is invariant under exchange \eqref{eq:interchange} in all cases. Therefore, the action $S_1''$ in \eqref{eq:S1''_nice} is simultaneously invariant under both background transformations \eqref{eq:k_transf} and \eqref{eq:k'_transf} and is also symmetric under exchange of the Killing vectors and background fields \eqref{eq:interchange}.

\section{Mixed 't Hooft anomalies between electric and magnetic symmetries}
\label{app:mixed_anomalies}

In this appendix, we derive the precise form of the mixed 't Hooft anomalies between the electric 1-form and magnetic $(d-3)$-form symmetries of the graviton theory.

\subsection{Four dimensions}
\label{app:mixed_anomalies_4d}

First, we consider the case of four dimensions, as studied in section~\ref{sec:anomalies_with_duals_4d}.
There, we derived an action $S_6^{d=4}$ which is invariant under \eqref{eq:electric_transf} and whose variation under \eqref{eq:B'_background_transf} is given by \eqref{eq:S6_var_4d_neater}. Recall that in four dimensions, the electric symmetries are labelled by Killing vectors, $k$, and the magnetic symmetries are labelled by $\A$- and $\C$-type CKY tensors, $K'$. We now consider the variation \eqref{eq:S6_var_4d_neater} for each choice of $k$ and $K'$ and show that it vanishes in some (but not all) cases. When it does not vanish and it is not possible to cancel the variation without spoiling invariance under \eqref{eq:electric_transf} then there is a mixed 't Hooft anomaly between the electric and magnetic symmetries.

\paragraph{Case 1: $k$ constant, $K'$ $\A$-type.}

If $k$ is a constant Killing vector, and $K'$ is an $\A$-type CKY tensor (i.e. a constant 2-form), then each term in the variation \eqref{eq:S6_var_4d_neater} vanishes. Therefore, the action $S_6^{d=4}$ given in \eqref{eq:S6_4d} is simultaneously invariant under both \eqref{eq:electric_transf} and \eqref{eq:B'_background_transf} in this simple case.

\paragraph{Case 2: $k$ not constant, $K'$ $\A$-type.}

The next simplest case to consider is when $K'$ is an $\A$-type CKY tensor (i.e. a constant 2-form) but $k$ is not a constant Killing vector (i.e. it is a rotational Killing vector in \eqref{eq:Khat_solution}). We parameterise them by
\begin{equation}
\begin{split}\label{eq:k_not_const_K'_Atype}
    k_\mu &= 2(d-3) \D_{\mu\nu} x^\nu \\
    K'_{\mu\nu} &= \A'_{\mu\nu}
\end{split}
\end{equation}
In this case, the variation \eqref{eq:S6_var_4d_neater} becomes
\begin{equation}\label{eq:S6_var_case2_working}
    \delta S_6^{d=4} = 6 \int \dd[4]{x} \delta B'_{\mu\nu} \delta^{\mu\nu\alpha\beta}_{\rho\sigma\gamma\delta} B_{\alpha\beta} {K'}^{\rho\sigma} \partial^\gamma k^\delta
\end{equation}
It follows from \eqref{eq:k_not_const_K'_Atype} that in this case $K'_{[\rho\sigma} \partial_\gamma k_{\delta]}$ is a constant 4-form. Therefore, in four dimensions it can be written
\begin{equation}\label{eq:K'dk}
    K'_{[\rho\sigma} \partial_\gamma k_{\delta]} = c_0 \epsilon_{\rho\sigma\gamma\delta}
\end{equation}
where
\begin{equation}
    c_0 \equiv -\frac{1}{4!} \epsilon_{\rho\sigma\gamma\delta} K'^{\rho\sigma} \partial^\gamma k^\delta = \frac{d-3}{12} \epsilon_{\rho\sigma\gamma\delta} \A'^{\rho\sigma} \D^{\gamma\delta}
\end{equation}
is a constant. Substituting \eqref{eq:K'dk} into \eqref{eq:S6_var_case2_working}, the variation can be written
\begin{equation}\label{eq:S6_var_case2_working2}
    \delta S_6^{d=4} = -24 c_0 \int \delta B' \wedge B
\end{equation}
which is of the form \eqref{eq:variation_general} given in the main text.

\paragraph{Case 3: $k$ constant, $K'$ $\C$-type.}

We now consider the case with a constant Killing vector $k$ but a $\C$-type CKY tensor $K'$. These are parameterised by
\begin{equation}
\begin{split}\label{eq:4d_case3_parameterisation}
    k_\mu &= -(d-3)\B_\mu \\
    K'_{\mu\nu} &= \C'_{\mu\nu\rho}x^\rho
\end{split}
\end{equation}
In this case, the variation \eqref{eq:S6_var_4d_neater} simplifies to
\begin{equation}\label{eq:S6_4d_var_case3_working}
    \delta S_6^{d=4} = -4 \int\dd[4]{x} \delta B'_{\mu\nu} \delta^{\mu\nu\alpha\beta}_{\rho\sigma\gamma\delta} B_{\alpha\beta} \tilde{K'}^{\rho\sigma\gamma} k^\delta
\end{equation}
In this case, the parameterisation \eqref{eq:4d_case3_parameterisation} implies that $\tilde{K'}^{[\rho\sigma\gamma} k^{\delta]}$ is a constant 4-form and so, in four dimensions, can be written
\begin{equation}
    \tilde{K'}^{[\rho\sigma\gamma} k^{\delta]} = c_1 \epsilon^{\rho\sigma\gamma\delta}
\end{equation}
where 
\begin{equation}
    c_1 \equiv - \frac{1}{4!} \epsilon_{\rho\sigma\gamma\delta} \tilde{K'}^{\rho\sigma\gamma} k^\delta = \frac{d-3}{4!} \epsilon_{\rho\sigma\gamma\delta} \C'^{\rho\sigma\gamma} \B^\delta
\end{equation}
is a constant. Inserting this in \eqref{eq:S6_4d_var_case3_working} means the variation can be written
\begin{equation}\label{eq:4d_case3_anomaly}
    \delta S_6^{d=4} = 16c_1 \int \delta B' \wedge B
\end{equation}
which is also of the form \eqref{eq:variation_general} given in the main text.

\paragraph{Case 4: $k$ not constant, $K'$ $\C$-type.}

The final case to consider is where $k$ is a non-constant Killing vector and $K'$ is a $\C$-type CKY tensor, parameterised by
\begin{equation}
\begin{split}\label{eq:4d_case4_parameterisation}
    k_\mu &= 2(d-3) \D_{\mu\nu}x^\nu \\
    K'_{\mu\nu} &= \C'_{\mu\nu\rho} x^\rho
\end{split}
\end{equation}
Using this, we find that the variation \eqref{eq:S6_var_4d_neater} is proportional to
\begin{align}
    \delta^{\mu\nu\alpha\beta}_{\rho\sigma\gamma\delta} \left( K'^{\rho\sigma} \partial^\gamma k^\delta - \frac{2}{3} \tilde{K}'^{\rho\sigma\gamma} k^\delta \right) &= - \frac{d-3}{3} \delta^{\mu\nu\alpha\beta}_{\rho\sigma\gamma\delta}  \left( 6 \C'^{\lambda\rho\sigma} \D^{\gamma\delta} + 4 \C'^{\rho\sigma\gamma} \D^{\delta \lambda} \right) x_\lambda \nonumber \\
    &= - \frac{10(d-3)}{3} \delta^{\mu\nu\alpha\beta}_{\rho\sigma\gamma\delta} \C'^{[\lambda\rho\sigma}\D^{\gamma\delta]} x_\lambda = 0
\end{align}
where the result vanishes as the 5-form $\C'^{[\lambda\rho\sigma}\D^{\gamma\delta]}$ vanishes identically in four dimensions. Therefore, in this case the variation vanishes 
\begin{equation}
    \delta S_6^{d=4} = 0
\end{equation}

In summary, there are mixed 't Hooft anomalies between several pairs of electric and magnetic 1-form symmetries of the four-dimensional graviton theory. The structure of these anomalies are summarised in Table~\ref{tab:4d_anomalies}.

\subsection{\texorpdfstring{$d>4$}{d>4} dimensions}
\label{app:mixed_anomalies_d>4}

We now consider the case of $d>4$ dimensions, as studied in section \ref{sec:anomalies_with_duals_d>4}. In that section, we derived an action $S_6^{d>4}$ given in \eqref{eq:S6_d>4_def} which is invariant under \eqref{eq:electric_transf} and whose variation under \eqref{eq:Btilde_background_transf} is given by \eqref{eq:S6_d>4_var}.

We will now demonstrate that this variation vanishes for some choices of Killing vector $k$ and CKY tensor $K''$, but not for others. Furthermore, when it does not vanish, it cannot be cancelled by a counter-term without spoiling invariance under the \eqref{eq:electric_transf} transformations.

\paragraph{Case 1: $k$ constant, $K''$ $\A$-type.}

If $k$ is a translational Killing vector (i.e. a constant vector) and $K''$ is an $\A$-type CKY tensor (i.e. a constant 2-form) then the variation \eqref{eq:S6_d>4_var} vanishes. Therefore, in this case the action $S_6^{d>4}$ is simultaneously invariant under both \eqref{eq:electric_transf} and \eqref{eq:Btilde_background_transf}.

\paragraph{Case 2: $k$ not constant, $K''$ $\A$-type.}

Next we consider the case where $k$ is a rotational Killing vector and $K''$ is still an $\A$-type CKY tensor. They are parameterised by
\begin{equation}
\begin{split}\label{eq:d>4_case2_parameterisation}
    k_\mu &= 2(d-3) \D_{\mu\nu} x^\nu \\
    K''_{\mu\nu} &= \A''_{\mu\nu}
\end{split}
\end{equation}
In this case, the variation \eqref{eq:S6_d>4_var} simplifies to
\begin{equation}
    \delta S_6^{d>4} = \frac{1}{2} \int \dd[d]{x} (\star \delta\tilde{B})^{\mu\nu} B_{\mu\nu} K''_{\alpha\beta} \partial^\alpha k^\beta
\end{equation}
From \eqref{eq:d>4_case2_parameterisation}, we note that 
\begin{equation}
    c_2 \equiv K''_{\alpha\beta} \partial^\alpha k^\beta = -2(d-3) \A''_{\alpha\beta} \D^{\alpha\beta} 
\end{equation}
is a constant. The variation can therefore be written
\begin{equation}\label{eq:d>4_case2_variation}
    \delta S_6^{d>4} = -c_2 \int \delta \tilde{B} \wedge B
\end{equation}
which is of the form \eqref{eq:variation_general} given in the main text.
This is non-vanishing and cannot be cancelled by a $d$-dimensional counter-term without spoiling invariance of the action under \eqref{eq:electric_transf}. This is nicely seen from the descent procedure, where the $(d+1)$-dimensional anomaly theory which reproduces this anomaly is
\begin{equation}\label{eq:d>4_case2_anomaly}
    -c_2 \int \tilde{B} \wedge \dd{B}
\end{equation}

\paragraph{Case 3: $k$ constant, $K''$ $\B$-type.}

We next consider the case where $k$ is a translational Killing vector, but $K''$ is now a $\B$-type CKY tensor. These are parameterised by
\begin{equation}
\begin{split}\label{eq:d>4_case3_parameterisation}
    k_\mu &= -(d-3)\B_\mu \\
    K''_{\mu\nu} &= \B''_{[\mu}x_{\nu]}
\end{split}
\end{equation}
and in this case the variation \eqref{eq:S6_d>4_var} simplifies to
\begin{equation}\label{eq:d>4_case3_var_working}
    \delta S_6^{d>4} = - \int \dd[d]{x} (\star \delta\tilde{B})^{\mu\nu} B_{\mu\nu} \hat{K}''_{\alpha} k^\alpha 
\end{equation}
We note that in this case, from \eqref{eq:d>4_case3_parameterisation},
\begin{equation}
    c_3 \equiv \hat{K}''_\alpha k^\alpha = \frac{d-3}{2} \B''_\mu \B^\mu
\end{equation}
is a constant. Therefore, the variation \eqref{eq:d>4_case3_var_working} can be written
\begin{equation}\label{eq:d>4_case3_variation}
    \delta S_6^{d>4} = 2c_3 \int \delta \tilde{B} \wedge B
\end{equation}
Again, this variation can be related to a BF-type $(d+1)$-dimensional anomaly theory similar to \eqref{eq:d>4_case2_anomaly}, 
\begin{equation}
    2c_3 \int \tilde{B} \wedge \dd{B}
\end{equation}
which is, again, of the form \eqref{eq:variation_general} given in the main text.
Therefore, there is no choice of $d$-dimensional counter-term which will cancel the variation \eqref{eq:d>4_case3_variation} without spoiling invariance under \eqref{eq:electric_transf}.

\paragraph{Case 4: $k$ not constant, $K''$ $\B$-type.}

Finally, we consider the case where $k$ is a rotational Killing vector and $K''$ is a $\B$-type CKY tensor. They are parameterised by
\begin{equation}
\begin{split}
    k_\mu &= 2(d-3) \D_{\mu\nu} x^\nu \\
    K''_{\mu\nu} &= \B''_{[\mu}x_{\nu]}
\end{split}
\end{equation}
Substituting these into the variation \eqref{eq:S6_d>4_var} gives a vanishing result. Therefore, the action $S_6^{d>4}$ is simultaneously invariant under both \eqref{eq:electric_transf} and \eqref{eq:Btilde_background_transf} in this case.

In summary, we have found a set of mixed 't Hooft anomalies between certain electric 1-form symmetries and magnetic $(d-3)$-form symmetries of the $d$-dimensional graviton. The cases where there are 't Hooft anomalies are shown in Table~\ref{tab:d>4_anomalies}.

\bibliographystyle{JHEP}
\bibliography{references}

\end{document}